\documentclass[sigconf,nonacm=true]{acmart}

\usepackage{graphicx}
\usepackage{color}
\usepackage{blindtext}
\usepackage{pifont}
\usepackage[useregional]{datetime2}
\usepackage{xspace}
\usepackage{adjustbox}
\usepackage{longtable}
\usepackage{array}
\usepackage{enumitem}
\usepackage{tikz}
\usepackage{amsmath}
\usepackage[english]{babel}
\usepackage{blindtext}
\usepackage{epigraph}
\usepackage[lined, ruled, linesnumbered]{algorithm2e}
\usepackage[justification=centering]{caption}
\usepackage{subcaption}
\usepackage[normalem]{ulem}
\useunder{\uline}{\ul}{}
\usepackage[font=small,skip=1ex]{caption}
\usepackage{needspace}

\usepackage{enumitem}
\setlist[itemize]{noitemsep, topsep=0pt}

\usepackage{wrapfig}

\renewcommand\footnotetextcopyrightpermission[1]{} % removes footnote with conference info
\definecolor{betterred}{RGB}{225,10,5}
\definecolor{gray}{RGB}{150,150,150}

\definecolor{green}{RGB}{40,225,40}
\definecolor{puce}{RGB}{204, 136, 153}
\definecolor{goldenrod}{RGB}{218, 165, 32}
\definecolor{purple}{RGB}{102, 0, 204}

\newcommand{\LATER}[1]{}
\newcommand{\LONGER}[1]{}
\newcommand{\deaddarling}[1]{}

\newcommand{\COMMENTOUT}[1]{}

\newcommand{\parab}[1]{\vspace{0.01in}\noindent{\bf #1}}

\newcommand{\srf}[1]{\S\ref{#1}}
\newcommand{\srfs}[2]{Sections~\ref{#1} and~\ref{#2}}
\newcommand{\frf}[1]{Fig.~\ref{#1}}

\newcommand{\trf}[1]{Table~\ref{#1}}

\newcommand{\ie}{\textit{i.e.,}\xspace}
\newcommand{\eg}{\textit{e.g.,}\xspace}

\newcounter{MYenumctr}

\newcommand{\highlightedit}[1]{\textcolor{black}{#1}}
\newcommand{\arxivedit}[1]{\textcolor{black}{#1}}

\newcommand{\AFM}{\textsf{AFM}\xspace}
\newcommand{\AFMs}{\textsf{AFMs}\xspace}
\newcommand{\NLM}{\textsf{NLM}\xspace}
\newcommand{\NLMs}{\textsf{NLMs}\xspace}

\newcommand{\Fathom}{\textsf{RPCAFMs}\xspace}

\begin{document}
\sloppy % prevent Latex from violating format rules

% \title{A \LaTeX\ Template for SIGCOMM 18}
\title{Do Data Center Network Metrics Predict Application-Facing Performance?}

%\titlenote{Produces the permission block, and copyright information}
%\subtitle{Extended Abstract}

% \author{Paper \#64, 13 pages body, 18 pages total including appendices}
\author{Brian Chang}
\affiliation{%
  \institution{University of Texas at Austin}
}
\email{bchang@cs.utexas.edu}

\author{Jeffrey C. Mogul}
\affiliation{%
  \institution{Google}
}
\email{mogul@google.com}

\author{Rui Wang}
\affiliation{%
  \institution{Google}
}
\email{ruiw@google.com}

\author{Mingyang Zhang}
\affiliation{%
  \institution{Google}
}
\email{mingyangzh@google.com}

\author{Aditya Akella}
\affiliation{%
  \institution{University of Texas at Austin}
}
\email{akella@cs.utexas.edu}

% Jeffrey C. Mogul (Google) <mogul@google.com>

% \renewcommand{\shortauthors}{X.et al.}

\begin{abstract}
Applications that run in large-scale data center networks (DCNs) rely on the DCN's ability to deliver application requests in a performant manner. DCNs expose a complex design and operational space, and network designers and operators care how different options along this space affect application performance. One might run controlled experiments and measure the corresponding application-facing performance, but  such experiments become progressively infeasible at a large scale, and simulations risk yielding inaccurate or incomplete results. 
Instead, we show that we can predict application-facing performance through more easily measured network metrics.  For example, network telemetry metrics  (e.g., link utilization) can predict application-facing metrics  (e.g., transfer latency). Through large-scale measurements of production networks, we study the correlation between the two types of metrics, and construct predictive, interpretable models that serve as a suggestive guideline to network designers and operators.
We show that no single network metric is universally the best predictor (even though some prior work has focused on a single predictor).  We found that simple linear models often have the lowest error, while queueing-based models are better in a few cases.

\end{abstract}

\maketitle
\pagestyle{plain}

\section{Introduction}~\label{sec:intro}
%The performance of distributed applications often depends on the characteristics of the data center networks (DCNs) where they run. 
%At scale, these 
Large-scale applications are distributed across multiple machines in multiple racks, and thus their performance
depends on behaviors, such as latency and throughput, of the data center networks (DCNs) where they run. 
These behaviors depend on multiple aspects of the underlying DCN, 
as well as the choices made by designers and operators.

We would like to evaluate how different aspects of network design and operation 
affect application performance. For example, when fabric planners are choosing
between DCN designs, or provisioning DCN capacity,
they want to know how their choices will affect application performance.
Network operators need to know whether a DCN is reaching an operating
point where application performance could be hurt.
Traffic-engineering optimizers need network-level  goals that preserve
application performance.
These use cases, and others, can depend on being able to \textbf{predict
application performance as a function of network behavior}. 

However, we cannot always make direct measurements of application performance, such as
web page load times or banking transactions per second.  
The reasons
include scale, data privacy, the intrusiveness of instrumenting applications
to measure top-level (e.g., user-facing) performance, and our lack of knowledge of future workloads.
Also, large-scale DCNs often support a complex mix of applications, which vary in
how they use network, compute, and I/O.

Instead, we can leverage metrics that indicate when
applications are bottlenecked by the DCN,  such as flow completion times (FCTs),
RPC latencies, and bulk-transfer goodput;
we call these \emph{application-facing performance metrics} (\AFMs).
We can then manage a DCN to preserve service-level objectives (SLOs) stated
in terms of \AFMs, under the reasonable assumption that this avoids
DCN-bottlenecked applications (but we could not validate that assumption in this study).

This viewpoint converts our problem of directly predicting application performance into one of predicting \AFMs from our knowledge of DCN design and workload characteristics, which is more practical but still difficult.

The gold-standard method to predict \AFMs is to run controlled experiments, and then directly
measure \AFMs such as flow completion times.
But those experiments are often infeasible: especially at scale,
it can be difficult to replicate realistic application mixes and workloads,
or even obtain the necessary hardware resources.
Alternatively, we could actively perturb the network to discover how this affects
a production workload, but the associated risks are usually unacceptable.

We could also run packet-level simulations based on application traces and record the simulated \AFMs
(e.g.,~\cite{ZhangNgEtAl2021,ZhaoEtAl2023}).
However, simulations cannot always produce accurate results:
First, simulators cannot always faithfully reflect the full complexity of a real DCN's dataplane behavior.
%can be too difficult for simulators to capture faithfully.
Second, realistic application traces are not always available, for reasons of scale or privacy.
Third, traces capture application behavior under a specific set of network conditions, and might not represent
behavior under different simulated conditions.

In this paper, we explore a different approach: we show how to create relatively
 \textbf{simple and intuitive predictive models that relate \AFMs to easily-measured \emph{network-level metrics}
(\NLMs)}.  Most production networks are already instrumented to collect \NLMs, 
such as link utilizations and discard rates, via mechanisms such as sFlow, SNMP, 
or OpenConfig.
Creating models that use \NLMs to predict \AFMs avoids the challenges of directly
collecting application-facing metrics through expensive (or infeasible)
experiments.

Using \AFM and \NLM traces for a large fraction of the applications in a set of production datacenters, we demonstrate that these models can indeed predict \AFMs in many cases, with
useful accuracy. They can also predict how operating a network in an overloaded
regime can lead to unpredictable and hard-to-bound \AFMs.

% \begin{wrapfigure}{r}{0.5\textwidth}
\begin{figure}[htb]
 \begin{minipage}{0.5\textwidth}
     \centering
     \includegraphics[width=.9\linewidth]{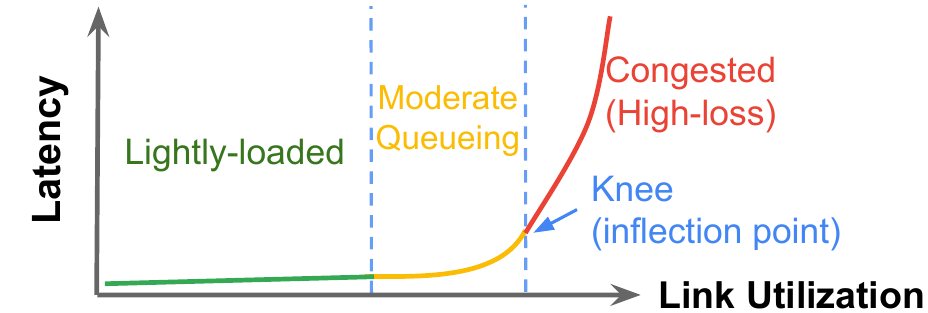}
     \caption{\small \emph{Hypothetical} latency vs. link utilization.}
     \label{fig:regimes-hypothetical}
 \end{minipage}
\end{figure}
% \end{wrapfigure}

We simplify the prediction challenge by considering network performance in three
regimes (Figure~\ref{fig:regimes-hypothetical}) : \emph{lightly-loaded}, where small changes in network metrics such as link utilization have little effect on \AFMs; \emph{moderate-queueing}, where
changes in \NLMs have significant but predictable impacts on \AFMs; and
\emph{congested}, where large queuing delays and/or packet losses lead to
severe and unpredictable application impacts.
Stakeholders want to avoid the congested regime, and manage their tradeoffs
within the other two regimes.

In our approach, we first model whether a network will be in the congested
regime, by \emph{detecting knees} in the curves representing \AFM vs. \NLM,
via a modified version of the Kneedle algorithm~\cite{kneedle}
(\srf{ssec:kneedle}).   Our goal for this is not to predict
application performance under severe congestion, but rather to predict
when congestion will set in, so that this ``danger zone'' can be avoided.

Once we have identified the knee (if any), we then try to make predictions
for the non-congested regimes.   
% For regimes where the \AFM vs. \NLM relationship is approximately linear, 
For these regimes, 
we create models using \textbf{quantile regression}~\cite{quantileregression} (\srf{ssec:quantile_regression},
which exposes sensitivity in both linear (lightly-loaded) and non-linear 
(moderate-queueing) regimes of a network, and is robust to noise and outliers in the measurements. 
Users can add \textbf{asymmetric bias} \srf{ssec:directional_loss}) to steer results towards overprediction or underprediction.

This paper makes the following contributions:
\begin{itemize}
    \item A method to produce trustworthy, actionable models that predict the relationships between
    a variety of \AFMs and \NLMs (\srf{sec:methodology}).  
    %Our method uses modified versions of the Kneedle algorithm
    %and quantile regression.
    This methodology is transferable and can be applied by
    DCN designers and operators using their own datasets.
    %We construct a comprehensive, interpretative methodology with operationally predictive power that captures the relationship between different network and application-facing metrics, including regression models and knee detection mechanisms. This methodology is transferable and can be applied to different DCNs, and can be used to analyze their own performance and fabric design principles.

    \item An in-depth case study (\srf{sec:eval_case_study}) on a real-world fabric, %along with fabric-specific operational insights.
    validated on data from multiple production fabrics (\srf{sec:multi-fabric-eval}).
    
    \item We identify which network metrics are good predictors, % of application-facing performance; 
    and show that \NLMs and \AFMs are correlated even in non-congested regimes.
    
    \item \highlightedit{
    We further validate our approach on data from multiple production fabrics (\srf{sec:multi-fabric-eval}).
    }
    
    \item \highlightedit{
    We empirically show that no single \NLM is the best predictor for the \AFMs studied (\srf{sec:best-nlms}).
    }
\end{itemize}
\section{Background and Motivation}
\subsection{Data center network topology}~\label{ssec:topo_overview}
Large-scale DCNs are typically designed to serve a diverse set of %customer-facing 
applications, with a goal of being agnostic to the application mix,
since this can change.
(Some DCNs are designed for specific large-scale applications, such
as high-performance computing or machine-learning training; these
single-function DCNs are outside the scope of this paper.)
% Examples of such applications include web servers, search engines, video streaming services, and more recently, machine learning workloads. 
%While different data centers contain various application mixes, one underlying goal when designing these data centers is to make them as generic as possible, since the application mix might change as time progresses.     
%To meet the goal of serving a wide variety of applications and optimizing for overall performance, spine-full Clos topologies are widely adapted due to its full-bisection bandwidth and capability of being constructed with cheap commodity switches. 

% \begin{wrapfigure}{r}{0.45\textwidth}
\begin{figure}[htb]
 \begin{minipage}{0.5\textwidth}
     \centering
     \includegraphics[width=.95\linewidth]{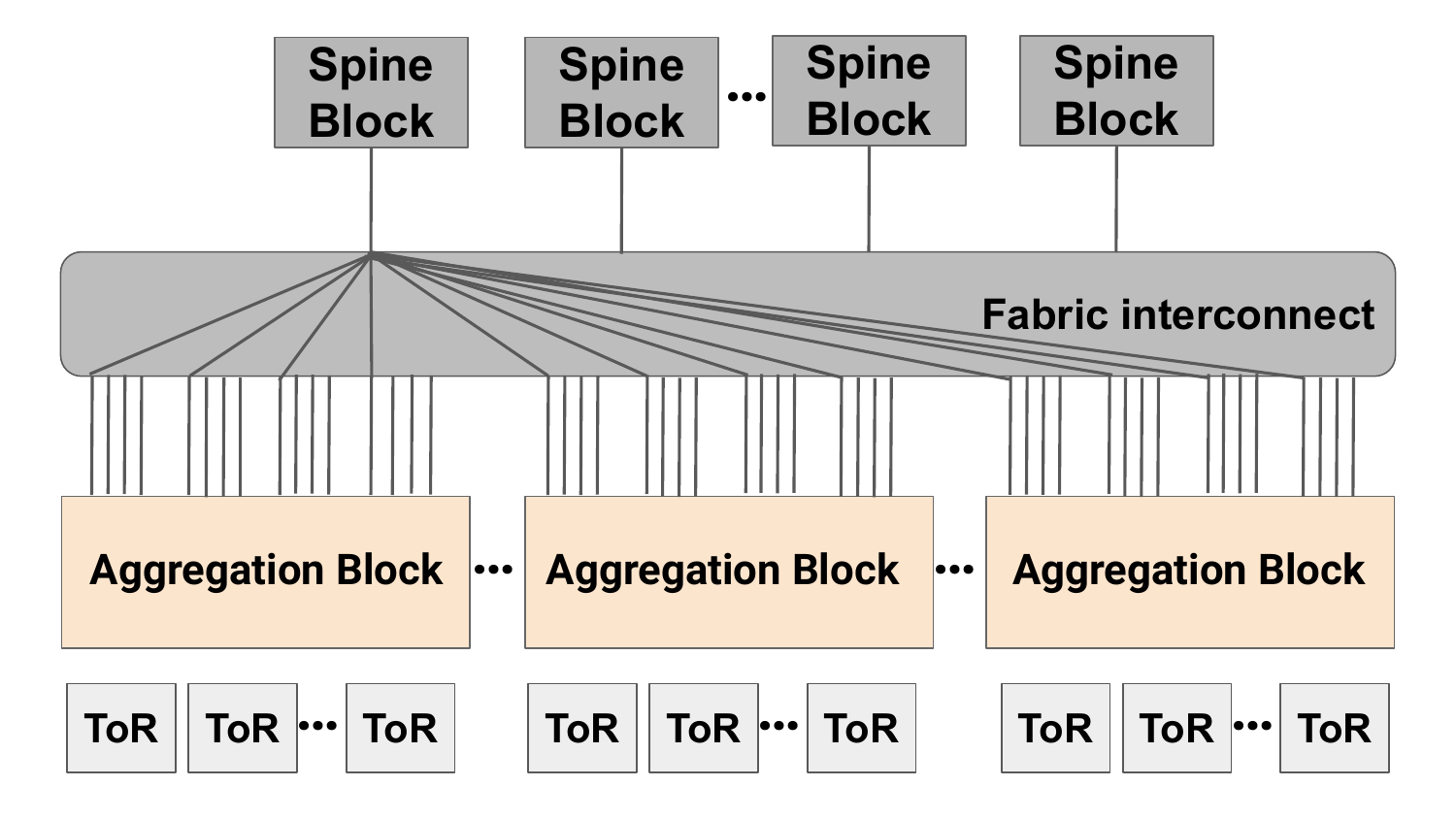}
     \caption{A typical Clos data center topology.}\label{fig:clos}
 \end{minipage}
\end{figure}
% \end{wrapfigure}

DCNs often rely on multi-path. multi-tiered topologies, such as Clos or Fat-Tree~\cite{fat-tree}, that
provide a good balance of cost vs. performance for these application mixes.
\frf{fig:clos} shows a typical example, with three layers of switches: 
%There are typically three layers of switches in such a topology: 
a Top-of-Rack (ToR) layer (to which machines are connected), an aggregation-block layer, and a spine-block layer. 
%ToR switches are directly connected to machines and have uplinks to the 
Aggregation and spine blocks are composed of several tiers of fully-connected switches. %The aggregation blocks are then connected to the spine blocks. 

While there are variations on this theme, such as Amazon's~\cite{CallaghanRivers2022},
the Facebook Fabric~\cite{fb-dcn},
or a spine-free Clos~\cite{jupiter-evolving}, the networks we study follow
the design in \frf{fig:clos}.

\subsection{Aspects of DCN performance}
\label{sec:related}
\subsubsection{Expected vs. Realistic Behavior}\label{ssec:expected_vs_realistic}
While different applications have different network communication patterns,
it has been shown that we can treat \AFMs as simple functions of \NLMs when the application is network-bottlenecked. For example, prior studies~\cite{ospf-weights,swift, ZhangEtAl2021}
have shown that high link utilizations lead to higher discard rates, which would presumably lead to higher flow completion times (FCTs).\footnote{However, an earlier work by Benson \textit{et al.} found no such correlation~\cite{BensonEtAl2010}.}

We might attempt to model the behavior of a network using queuing theory, but the complexities
of real-network behavior does not always conform to simple models.
At moderate utilizations, our results (\srfs{sec:eval_case_study}{sec:multi-fabric-eval}) 
are indeed consistent with modeling the DCN as a single, giant M/D/1 queue, 
which implies a reciprocal relationship between latency and utilization~\cite{qt}.
However, when the network is lightly loaded, with little or no persistent queueing,
we often see a linear relationship between \AFMs and \NLMs;
this is especially true for high-priority QoS (\srf{ssec:qos}) classes that get preferential treatment.
At high utilizations, packet losses due to full queues make queueing-theory models much harder to apply.
% Even at moderate utilization, Clos networks may have multiple queues, with different behaviors.
Our goal is therefore to help stakeholders distinguish between these three operating regimes, and
provide them with regime-specific predictive models, so that they
can take appropriate actions.

\subsubsection{Quality-of-service for Datacenter Applications}~\label{ssec:qos}
Large-scale DCNs often allow some applications to select Quality of Service (QoS) classes, giving their packets preferential
treatment, to protect latency, loss, or throughput SLOs when link bandwidth or switch buffers are scarce.
In this paper, we focus on three QoS classes:

\begin{itemize}
    \item \textbf{High priority}:  for applications that require low latency, and very low loss.
    \item \textbf{Medium priority}:  for applications that accept some latency and a limited level of loss.
    \item \textbf{Low priority}: for applications that can tolerate more latency and/or more loss.
\end{itemize}

\subsection{Application performance stakeholders}~\label{ssec:stakeholders}
Various parties and systems would benefit from a deeper understanding of the relationship between \NLMs and \AFMs
(the \textbf{\NLM-\AFM relationship}).
These stakeholders are:\\

\parab{Fabric designers}, who analyze competing DCN designs
(e.g., \cite{fat-tree,Namyar2021,jupiter-evolving,FlattenedClos})
to estimate their impacts on application performance. 
\highlightedit{
For example, designers often use forecasts of \NLMs, such as expected utilization, to ensure sufficient capacity is built to serve the predicted demand~\cite{Gersteletal2014, Hangetal2021}. 
}
Data on the \NLM-\AFM relationship gives a designer insights into which candidate designs are likely to support better application performance.
E.g., if the switch hardware in an aggregation block is currently underprovisioned, leading to high utilization and reduced application performance, 
a designer can extrapolate from the \NLM-\AFM relationship to plan a cost-effective hardware upgrade that meets application needs.  This might involve adding uplinks, 
or upgrading to switches with higher port speeds.

\parab{DCN operators}, who would like to avoid application SLO violations based on definitions of different QoS classes and objectives. 
However, it is difficult to directly instrument the underlying applications due to privacy and scale issues. 
Understanding the \NLM-\AFM relationship helps operators detect signs of current
or impending
SLO violations, and allows them to implement corresponding mitigations, such as workload steering in the congested aggregation blocks.

\parab{Automated traffic and topology engineering systems}
that rely on network-level measurements to make reconfiguration decisions (e.g., ~\cite{HeEtAl2007,KandulaEtAl2005,WangEtAl2006}). 
\highlightedit{
Traffic engineering (TE) is commonly used by large service providers to optimize traffic flowing through data centers and wide-area networks. State-of-the-art TE systems solve an optimization problem that decides how to split traffic through multiple paths for traffic demand between a given source and destination. The objective for this problem is typically a function derived from an \NLM, with \textit{maximum link utilization} (MLU) being the most commonly-used metric~\cite{mitraEtAl1999,applegateEtAl2004,agarwalEtAl2005,WangEtAl2006,jupiter-evolving}.
However, it is unclear how well MLU correlates to \AFM performance, or whether MLU is the ideal \NLM to optimize TE.
}

%\jeff{explain what "Topology Engineering" is; I don't think anyone outside Google uses this term}
\highlightedit{
Topology engineering (ToE), which dynamically reconfigures capacity between aggregation block pairs to better match a non-uniform traffic demand, follows a similar form, where an optimization problem with an \NLM-based objective is used to decide how to assign capacity between block pairs~\cite{jupiter-evolving}. 
}
Understanding the \NLM-\AFM relationship can help designers of
these systems potentially choose optimization objectives that better reflect application-facing performance.
For example, if a topology engineering system observes a network metric that exceeds a pre-defined threshold, and deduces that applications are at risk of SLO violations, the system can increase the network capacity between certain aggregation-block pairs.

\parab{Job schedulers}, when choosing when to admit or place workloads,
can leverage the \NLM-\AFM relationship to control rates at which hosts can send traffic into the DCN. For example, schedulers can place tasks or VMs to
minimize congestion~\cite{LiEtAl2018}, 
or temporarily throttle lower-priority flows to
preserve high-QoS SLOs~\cite{KumarEtAl2015,PerryEtAl2017}.

\parab{Network researchers}, who
would like to know whether new network designs can better serve applications. It is difficult for academic researchers to gain access to large-scale workload traces and infrastructure that would allow them to directly measure application-facing performance. Knowing
realistic \NLM-\AFM relationships would help researchers extrapolate application-facing behavior through network simulation,
to validate their results.

\subsection{Requirements of predictive models}
\label{sec:modelreqs}

\epigraph{All models are wrong, but some are useful.}{George Box}

Our models must be both \textbf{actionable} and as \textbf{trustworthy} as possible.

%First, models need to produce 
Actionable results allow stakeholders, such as designers and operators, to take
appropriate measures to ensure application performance. 
To be actionable, model-based outputs must be \emph{interpretable} in terms of application
behavior, so that stakeholders can
make straightforward decisions about network design and operation.
We also need to \emph{differentiate between overprediction and underprediction} of \AFMs,
because stakeholders generally need to bias these errors in one direction -- when balancing
application safety (SLO compliance) against infrastructure costs, most users treat SLO violations
as the greater concern.

Stakeholders cannot rely on models that they do not trust. 
In particular, models need to be (1) \emph{robust against outliers}, since measurements are usually noisy,
%to be sure models don't break under noisy measurement data, 
and (2) provide \emph{confidence or error bounds}, allowing users to understand the limits of trustworthiness,
since most models are only reliable within certain ranges.

Our goal, therefore, is to provide models that accurately predict \AFMs from \NLMs when we have high confidence in those predictions,
and to inform the user when we cannot provide a high-confidence model.

\highlightedit{
Some modeling methods fail to meet 
%Not all types of methods meet 
our requirements.
For example, while machine learning has demonstrated robust predictive powers over various complex tasks~\cite{Hangetal2021, mao2019learning, valadarsky2017learningtoroute},
these black-box models do not provide interpretable predictions.
Traditional regression methods are highly sensitive to outliers.
High-fidelity simulations are not generally feasible at DCN scales.
Real-world experiments, 
without some sort of model,
%on their own,
are neither feasible nor predictive;
%Without some sort of model, 
an experiment can only tell us what happens under the tested conditions, not what might happen with a different network design or configuration. Table~\ref{tab:model_choices} summarizes these approaches.
}

\begin{table*}[]
% \begin{adjustbox}{width=\textwidth}
\small
\begin{tabular}{|l|l|l|l|l|l}
\hline
& \multicolumn{4}{c|}{Requirement} \\
\hline
Network Modeling Approach                & Interpretability & Robustness
& Error bounds & Realistic Scenarios \\ \hline \hline
Deep Neural Networks      & $\times$                & \checkmark                  & \checkmark            &  \checkmark                                          \\ \hline
% Generalised Additive Models & $\triangle$         & $\triangle$                     & \checkmark                        & \checkmark                                          \\ \hline
Regression                & \checkmark         & $\times$                  & \checkmark                               & \checkmark                                          \\ \hline
High-fidelity Simulations & $\triangle$         & N/A                & N/A                                         & $\times$                                          \\ \hline
Quantile Regression       & \checkmark                & \checkmark                  & \checkmark            & \checkmark
\\\hline
\end{tabular}
% \end{adjustbox}
\\{Requirement satisfaction: \checkmark: full; $\triangle$: limited; $\times$: not met; N/A: not applicable for the approach.}
\caption{Comparison of \NLM-\AFM prediction approaches.}
\label{tab:model_choices}
\end{table*}

\section{Methodology Overview}
\label{sec:overview}

\highlightedit{
We now introduce a methodology that meets all the requirements above, by combing \textit{quantile regression} and \textit{knee detection}. Quantile regression simultaneously provides interpretability, robustness, error bounds, and also covers tail behavior and realistic scenarios. Knee detection provides an easily interpreted signal for when a network enters a highly-congested regime.
}
% This section briefly summarizes our approach to generating predictive models.  
This section briefly summarizes our approach.
Subsequent sections develop each aspect in greater detail.

\begin{figure*}[htb]
     \centering
     \includegraphics[width=\linewidth]{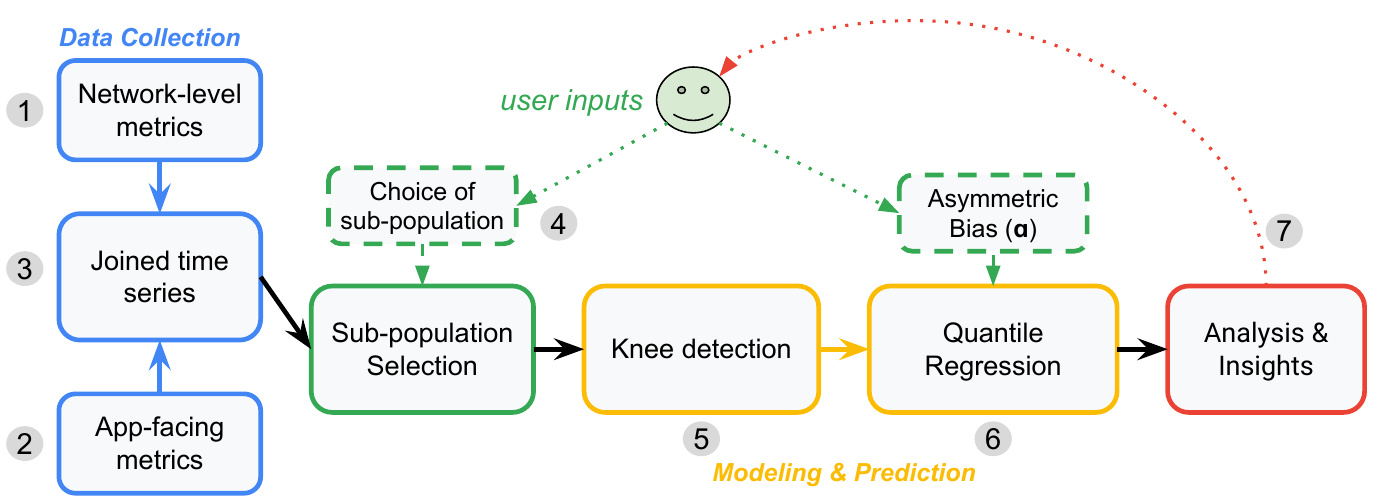}
     \caption{Overview of the approach}
     \label{fig:overview}
\end{figure*}

We start by gathering data for each DCN fabric of interest (\srf{sec:fabric_choice}), which allows us to
make fabric-specific predictions.  We currently treat each fabric as an independent
prediction problem, because each fabric may have a unique mix of applications.\footnote{
One possible topic for future work is to decide if two fabrics $A$ and $B$ are sufficiently
similar that a model based on fabric $A$ can be used to predict the behavior of
fabric $B$; we expect this would work in many cases, but we have not tested that hypothesis.}
For each fabric, we collect two datasets: time series of \NLM metrics (\srf{sec:nlm-telemetry},
\textcircled{\small{1}} in \frf{fig:overview}) and
corresponding time series of \AFM metrics (\srf{sec:rpc-data}, \textcircled{\small{2}}).
The \NLM series come from standard
switch-based network-monitoring sources, and cover all data-plane switches in these fabrics.
The \AFM series come from RPC-level instrumentation, and
so these only cover the subset of applications which both use RPCs and allow us to instrument them.
Aside from that limitation, the \AFM traces are agnostic to application type.

% \begin{wrapfigure}{r}{0.5\textwidth}
\begin{figure}[H]
 \begin{minipage}{0.5\textwidth}
     \centering
     \small
     \includegraphics[width=.95\linewidth]{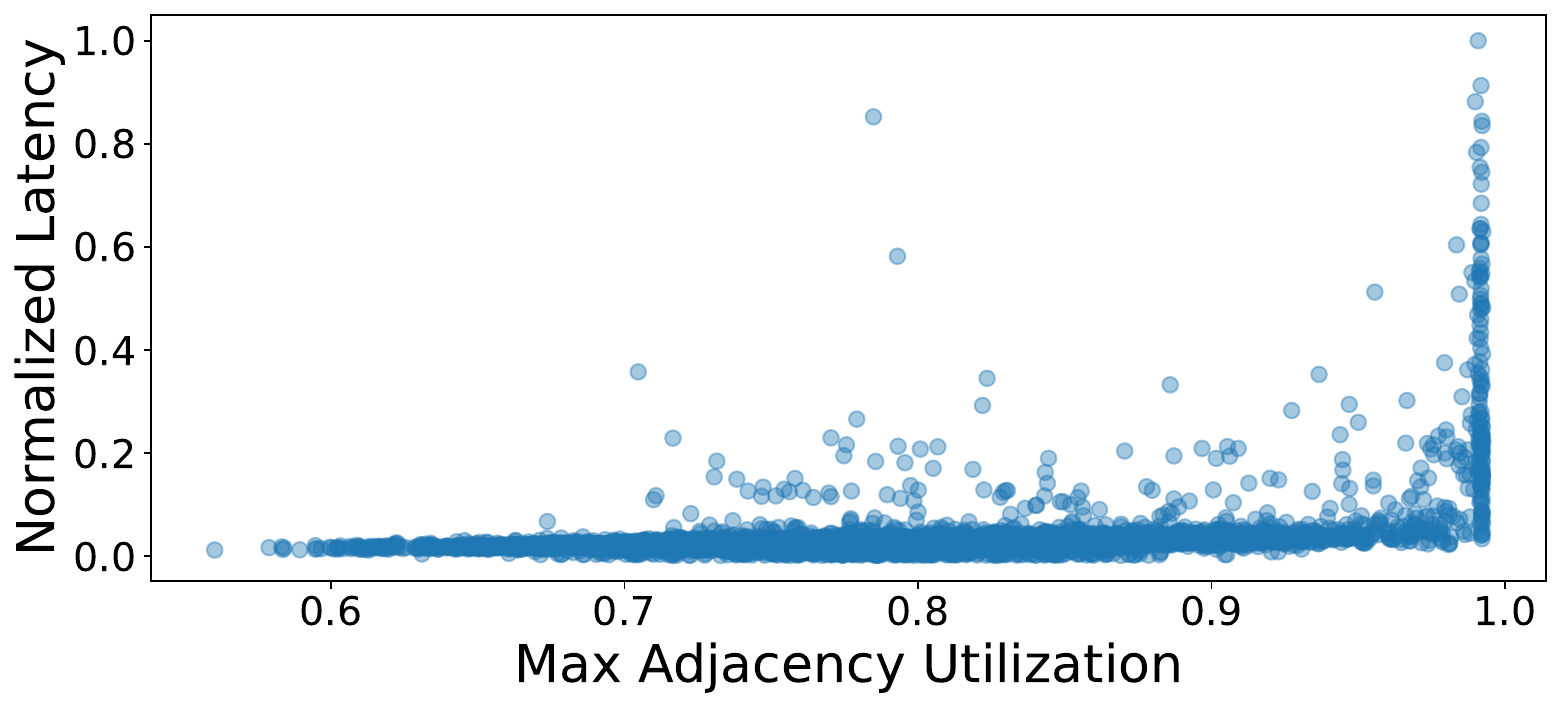}
    %  \\ \fixthis{We will replace this with a normalized y-axis figure}
    \\ Max Adjacency Utilization is defined in Table~\ref{table:netdata}
    \caption{Example scatterplot of latency vs. utilization}
     \label{fig:scatter-plot-example}
 \end{minipage}
\end{figure}
% \end{wrapfigure}

Then, using the joined time-series datasets \textcircled{\small{3}}
for each fabric, the user may select a sub-population \textcircled{\small{4}}, 
such as the data for a specific QoS.
We then analyze a number of pairwise data relationships between
\NLMs (such as link utilizations or discard rates) and \AFMs (such as RPC latencies at different message sizes).
It is helpful to visualize these pairs as scatterplots, with the \NLM on the x-axis and the \AFM
on the y-axis; an example is shown in \frf{fig:scatter-plot-example}.
For each such pair, we start by looking for ``knees'' -- significant points of inflection --  in the curve,
using a modified version of the previously-published Kneedle algorithm 
(\srf{ssec:kneedle}, \textcircled{\small{5}}).
We assume these knees, if any, represent the transition to a congested regime; these become our predictions
about where that transition will occur.

We then treat the data points to the left of a knee (if any) as representing the uncongested
behavior of the fabric, and apply \emph{quantile regression} to that subset of the data
(\srf{ssec:quantile_regression}, \textcircled{\small{6}}).   
Quantile regression is insensitive to outliers, and naturally provides confidence bounds for each prediction.
Our approach to quantile regression supports both linear (lightly-loaded) and non-linear
(moderate-queueing) behaviors, and how to distinguish between these regimes.
It also supports user-tunable asymmetric bias, to favor SLO preservation over cost
reduction (or vice versa).

\parab{Practical implications}
Network designers and operators, including automated control planes such as for
traffic engineering, generally want to avoid congestion.   Therefore, our ability
to detect knees in the \NLM-\AFM relationship, if any, help to avoid these danger
zones.  For non-congested regimes, the results of quantile regression help designers
and operators understand how application-facing performance might be affected by changes
in \NLMs \textcircled{\small{7}}.

When we have high confidence in a linear relationship, the slope provides a simple
summary of the \NLM-\AFM relationship.  When instead we have a high-confidence
fit to a queueing-theory model, such as M/D/1, this suggests that relationship
is reciprocal.

\section{Data Collection Pipeline}
\label{sec:dataset}

This section describes the data we collected, both for developing our approach and for
evaluating it.

\subsection{Choice of fabrics}
\label{sec:fabric_choice}
We collected \NLM and \AFM data for 19 different fabrics over a span of 2 months. 
These fabrics all use a Clos-like topology (\frf{fig:clos}) and serve generic applications.
We choose the fabrics so that the value ranges of \NLMs and \AFMs we collect are as diverse as possible, allowing
us to test our methodology over multiple regimes (lightly-loaded, moderate-queueing, and congested).

\highlightedit{
Note, however, that there is an inherent paradox in this dataset: our data
% is notable precisely because it 
comes from production workloads on production
networks, but most production datacenter networks are carefully managed to
avoid becoming performance bottlenecks.}
In particular, datacenter operators seldom allow significant congestion to persist, so we only have a
few examples of highly-congested fabrics (\ie where operators have determined that this is acceptable to users).

\subsection{Network telemetry pipeline}
\label{sec:nlm-telemetry}

In the fabrics we studied, the existing network-telemetry pipeline collects 
port-level
switch statistics at
30-second intervals.  This data covers switches at all layers (ToRs, aggregation blocks, and spine blocks)
using a vendor-neutral industry-standard data model.   Most enterprises have similar telemetry pipelines.
\trf{table:netdata} lists the metrics we considered in our analyses, including several metrics that
we derive in a simple post-collection pass.

\begin{table*}[htb]
\centering
\caption{Port-,  link-, and adjacency-level metrics provided by our network telemetry pipeline.}
\small
\begin{adjustbox}{width=\textwidth}

\begin{tabular}{|l|c|l|} 
 \hline
 Name & Data Type & Description \\ \hline\hline
 Fabric Name & Metadata & Name of the fabric that contains the switch port. \\  \hline
 Stage & Metadata & Layer (i.e. ToR, aggregation, spine) where the switch is located. \\ \hline
 Port ID & Metadata & Port IDs of the switch and the peer switch. \\  \hline
 Port Speed & Metadata & Measured in bits per second. \\  \hline
 Src/Dst Aggregation Block & Metadata & Source and destination aggregation block containing the port. \\  \hline
%  Total Discards & Metric & Number of packets that were discarded at the input/output port. \\  \hline
%  Total Packets & Metric & Number of packets that were received to/sent from the port.  \\ \hline
 Total Octets & Metric & Sum of number of bytes that were received/sent at the port.  \\ \hline
\arxivedit{Incoming Octets} & \arxivedit{Metric} & \arxivedit{Number of bytes that were received at the port.}  \\ \hline
\arxivedit{Outgoing Octets} & \arxivedit{Metric} & \arxivedit{Number of bytes that were sent at the port.} \\ \hline
 Link Utilization & Derived & Sum of outgoing bits divided by port speed. \\ \hline
 Max/Average Link Utilization & Derived & Max/Average link utilization across all links in the fabric. \\ \hline
%  Average Link Utilization & Derived & Average link utilization across all links in the fabric. \\ \hline
 k-th Percentile Link Utilization & Derived & The k-th percentile link utilization across all links in the fabric. \\ 
 \hline
 P5-P95 distance: Link Util. & Derived & The difference between the 5th and 95th percentile link utlization. \\ \hline
 Adjacency Utilization & Derived & Sum of outgoing bits divided by port speed. \\ \hline
 Max/Average Adjacency Utilization & Derived & Max/Average adjacency utilization across all links in the fabric. \\ \hline
%  Average Link Utilization & Derived & Average link utilization across all links in the fabric. \\ \hline
 k-th Percentile Adjacency Utilization & Derived & The k-th percentile adjacency utilization across all links in the fabric. \\ 
 \hline
 P5-P95 distance: Adjacency Util. & Derived & The difference between the 5th and 95th percentile adjacency utlization. \\ \hline

\end{tabular}
\end{adjustbox}
%\caption{Data format of our network telemetry pipeline.}
\label{table:netdata}   
\end{table*}

By exploiting metadata, such as the prefixes of chassis names, we have the flexibility to group the network data
at various granularities.   We can look at individual link-level statistics -- \eg basing a prediction
on the most-utilized link between a pair of blocks -- or we can analyze aggregations
of links, formally defined as \textbf{adjacencies} -- \eg treating all links between a pair of blocks as a single aggregated link.
Coarse-grained, adjacency-level utilization exposes fabric-wide behavior, while fine-grained, port-level link utilization 
exposes outlier behaviors within the fabric.  

\subsection{RPC instrumentation}
\label{sec:rpc-data}

A large fraction of the traffic in the fabrics we studied comes from RPC-based applications.  The RPC layer used in
most of these applications includes transport-level instrumentation, \Fathom (similar to Fathom~\cite{Fathom2023}), that passively collects some \AFMs.

\Fathom breaks down each RPC into a set of events,
%We make use of our ability to monitor production transport-layer performance of RPCs through the kernel TCP/IP stack. The instrumentation system is a passive monitoring system that breaks down RPCs into events, 
each associated with high-resolution kernel timestamps as the RPC requests and responses traverse the stack.
This allows \Fathom to collect both message-oriented metrics, such as RPC transmit latencies,
and connection-oriented metrics, such as delivery rates (bytes/sec).
\trf{table:appdata} lists the metrics that we use in this paper.

%The instrumentation system primarily reports three types of metrics: transfer latencies and delivery rates, which are message-oriented; and connection states (e.g. Min RTT, retransmission), which are connection-oriented. Some of these metrics and their metadata, along with their brief descriptions, are summarized in Table~\ref{table:appdata}. 
% Due to the large amount of RPCs issued, our system uses sampling in order to keep the instrumentation scalable. 

\begin{table*}[htb]
\centering
\caption{RPC metrics collected}
\begin{adjustbox}{width=\textwidth}
\begin{tabular}{|l|c|l|} 
 \hline
 Name & Data Type & Description \\ \hline\hline
 Transmit Latency & Message-Oriented Metric & Time from when the first byte is sent to a NIC to when an ACK is received for a prefix of the message. \\  
 && Values are separately measured for small, medium and large RPCs. \\\hline
 Delivery Rate & Connection-Oriented Metric & Non-application-limited throughput of the connection. \\\hline
%  && Values separately measured for small and large RPCs.\\ \hline
%Min RTT & Connection-Oriented Metric & The TCP connection’s estimation of the minimum Round Trip Time (RTT).  \\ \hline
 Traffic Class & Metadata & The traffic class (i.e. QoS class) of the issued RPC. \\  \hline
 Congestion Control & Metadata & The underlying congestion control algorithm used in the TCP connection. \\  \hline
 Src/Dst Aggr. Block & Metadata & The source and destination aggregation blocks of the RPC request. \\  \hline
 %Src/Dst Job & Metadata & The source and destination jobs of the RPC request. \\  \hline
 %Src/Dst User & Metadata & The source and destination users of the RPC request. \\ \hline
%  Retransmissions & Metric & The ratio of total bits/packets that TCP has sent for more than one time. \\ \hline
\end{tabular}
\end{adjustbox}
%\caption{Data format of our RPC instrumentation system.}
\label{table:appdata}
\end{table*}

To avoid the computational and storage demands from saving raw data on billions of RPCs issued daily in our fabrics,
%and to reduce interference with application performance,
\Fathom leverages \textbf{t-digests}~\cite{tdigest}.
T-digests preserve
%to limit compute and storage resource usages, while maximally preserving 
the original distribution of the data through \textbf{approximated quantiles}. Compared to other compression methods, such as binning with histograms, the t-digest data structure has high approximation accuracy, especially at the tail, and can be merged with other independent t-digests (e.g. through common SQL operations like “GROUP BY”). Based on the calculated t-digests, we summarize the RPC metrics into different percentiles of the metrics’ distribution. We can also group the derived t-digests based on different \AFM attributes to support a wide range of analyses.

To further reduce overhead, \Fathom reports data aggregated over 5-minute intervals, so when we join this data
with \NLM data, we also re-aggregate the \NLM data over 5-minute intervals.  We would prefer to use shorter
intervals, but this tradeoff is currently necessary for feasibility.

Note that, because \Fathom was designed for a different purpose and to execute with minimal impact on applications, the way it
reports data grouped by RPC sizes is a little unusual.  For example, for message lengths between 1 KiB and 8 KiB,
\Fathom reports transmit latency from when the first byte is handed to the NIC until TCP receives the acknowledgement
for the last byte of the \emph{first 1 KiB} of the message.  (See \frf{fig:fathom-timeline}.)
Similarly, for messages between 64KiB and 256KiB, latency is measured for the first 64KiB.
%\aditya{what does it do for messages lager than 8K?}
This is therefore (1) not the same thing as the round-trip \emph{RPC latency}, since it measures each direction of an RPC independently, and
(2) only measures transmit latency for a prefix of the message.
However, we believe that \Fathom-based \AFMs are a reasonable proxy for actual application-facing network performance.

Enterprises without a system like \Fathom might instead use mechanisms such as DTrace~\cite{CantrillEtAl2004} to collect \AFMs.
\section{Methodology and algorithms}
\label{sec:methodology}

Our methodology incorporates two components:  knee detection with Kneedle, and quantile regression.
The Kneedle algorithm, based on geometry theory, detects whether \AFMs experience congestion after a certain \NLM threshold.
Quantile regression provides statistical predictions of the \NLM-\AFM relationship, with strong noise/outlier resistance. 
 
% The following subsections describe these components in detail.  

\subsection{Knee Detection with Kneedle}
~\label{ssec:kneedle}
The ``knee'' of a curve is a point where the slope of a curve changes suddenly -- \ie has a high second derivative.
In the analysis of computer system performance, a knee in the relationship between an independent variable
(\eg an \NLM) and a dependent variable (\eg an \AFM) often represents a transition from ``good'' to ``bad''
behavior.  In packet networks, this typically represents a transition to a congested regime.
Therefore, we would like to predict whether an \AFM-vs-\NLM relationship has knee, and if so, where it is.

%In computer systems, there are often turning points where the cost of increasing the magnitude/capacity of some adjustable parameter starts receiving a diminishing magnitude of performance improvement; or where the system cannot tolerate additional increase in certain overheads without severely degrading performance. These turning points are commonly referred to as “knees” by researchers, and represent operating points with substantial behavioral change on different sides.

One challenge is that the definition of a ``knee'' is informal and heuristic, and can vary dramatically
depending on application-specific objectives and the chosen coordinate scales.   
We believe the concept is still useful, since it aligns well with an intuitive explanation of network behavior -- for
example, the rapid increase in RPC tail latency that occurs above a threshold network utilization.

For knee detection, we use the previously-published Kneedle algorithm~\cite{kneedle}.
%One challenge in knee detection is that knees are mostly heuristically defined: the definition of a knee differs drastically among prior literature depending on the their application-specific objectives.
%To tackle this problem, Kneedle’s uses 
Kneedle formalizes the intuitive notion of knee as the point of maximum curvature
in a function (\ie its deviation from a straight line).
A potential knee is defined as the $x$ value where there is local-maximum curvature.

Kneedle operates on a curve, not on a set of (\NLM, \AFM) samples, so we define the envelope of
a dataset as the 95th percentile \AFM value in each of $k$ buckets across the dataset.
(This captures the bulk of the dataset without being too sensitive to outliers.)
We chose $k$ to match the number of buckets appropriate to the x-axis extent of the data; see
\srf{ssec:setup} for how we do that.
These choices
are somewhat arbitrary, but we found that algorithm finds almost the same knees no matter what percentile
we used.
Because knee-detection is inherently use-case-specific, we modified the Kneedle algorithm's criteria
for finding a knee in two ways:
\begin{enumerate}
    \item \textbf{Curvature of the potential knee is larger or equal to threshold $\mathcal{C}$.} Not all \AFM-\NLM metrics have interesting knees;
    we only want to find points that represent drastic degradation in the \AFM. Therefore, we enforce a minimal threshold $\mathcal{C}$ for the curvature of a detected knee.
    In comparison, vanilla Kneedle returns the point of maximum curvature as the knee, regardless of its absolute value.
    $\mathcal{C}$ is adjustable based on the risk tolerance of operators. 
    \item \textbf{Curvature of the detected knee is a global maximum.}  We would like to avoid any local maxima, which can result from
    minor measurement noise, and so we only return the
    worst (maximum-curvature) knee.
\end{enumerate}

The Kneedle algorithm also requires users to specify the convexity/concavity of the relationship, and assumes that the relationship is strictly increasing or decreasing. 
% While we expect some behavioral deviations of the \NLM-\AFM relationship from the expected behavior (\srf{ssec:expected_vs_realistic}), 
% and measurement noise also causes some local deviations,
% we generally do see a reliably convex relationship between \NLMs and latency-based \AFMs.
\arxivedit{While we expect some behavioral deviations from the ideal \NLM-\AFM relationship (\srf{ssec:expected_vs_realistic}) due to measurement noise and other potential factors, we generally observe a reliably convex relationship between \NLMs and latency-based \AFMs.}
\arxivedit{This pattern emerges as the data center network behaves like a large-scale queueing system, formed by collective buffers of network switches. Queueing theory predicts that delay in such systems follows a convex relationship.} 
Conversely, we assume a concave relationship between \NLMs and throughput-based \AFMs, \arxivedit{where we expect per-application/RPC throughput to decrease as the network becomes increasingly saturated}.

In a few cases, Kneedle finds a false knee, either much smaller than the ``true'' knee, or when there is no clear knee at all.
We might be able to improve the algorithm to avoid that; this is future work.

\subsection{Quantile Regression}
~\label{ssec:quantile_regression}
For \NLM-\AFM relationships to the left of a knee, or where there is no clear knee, we use
regression analysis to provide a simple predictive model for \AFMs.
For this, we use \emph{quantile regression}~\cite{quantileregression}, a type of regression analysis that estimates the \textbf{quantiles} (e.g. the median, the 90th percentile, etc.) of a target variable.
This differs from classical regression models that directly predict the target variable.

Quantile regression provides two advantages over classical regression models:
(1) quantile regression is extremely robust to noise, because a small number of outlier values have minimal effect on the calculated quantiles; 
and (2) with quantile regression, we can inspect different quantiles of the same dataset,
to obtain a more well-rounded view of the data (\eg by looking at both median and tail percentiles of a target variable).
We can apply quantile regression both for linear models (\eg for lightly-loaded network regimes), and for non-linear models
(\eg for moderate-queuing regimes); we discuss the latter in \srf{sssec:feature_transformation}.

To provide a brief overview of quantile regression’s formulation, we use a linear model as an example.
Consider a classical linear regression formulation,
with a single predictor variable ($x$) and a single target variable ($y$): 
\begin{equation}\label{eq:linreg}
y= \beta x+c
\end{equation}
$\beta$ is known as the \textit{slope}, and $c$ is known as the \textit{intercept} of the equation.
We can interpret Eq.~\ref{eq:linreg} as the expected change in y per unit change in x. 
In linear quantile regression, instead of estimating the expected change in the y value, we 
estimate the $\tau$-th percentile of $y$, where $\tau$ ($0\leq\tau\leq1$) is defined as the probability that the targert variable $y$ is smaller or equal to $\mu_{\tau}$, the $\tau$-th percentile of $y$.
\begin{equation}
    \tau = Pr(y\leq\mu_{\tau})\equiv F_{y}(\mu_{\tau})
\end{equation}
\begin{equation}
    \mu_{\tau} = F_{y}^{-1}(\tau)
\end{equation}
We can then define the conditional quantile of $y$ as:
\begin{equation}\label{eq:conditional_quantile}
    y_{\tau|x}=\mu_{\tau|x}=F_{y|x}^{-1}(\tau|x)=\beta x + c 
\end{equation}
where conditioning on $x$ produces $(x, y_{\tau})$ pairs that can then be modeled using common regression techniques.

\subsubsection{Data Discretization}
As shown in Equation~\ref{eq:conditional_quantile}, $y_{\tau|x}$ conditions quantile values of $y$ on certain values of $x$. 
If we have the ground-truth distribution of the underlying $x, y$ relationship, we can directly calculate $\mu_{\tau}(x)$ through its PDF. 
However, we seldom have the ground truth of the actual distribution. 
Instead, we typically have a series of $(x, y)$ data points, which are then used to estimate the ground truth of the distribution. 
In order to approximate the conditional quantiles, in quantile regression the predictor variable $x$ is first \textbf{bucketed} into equal-sized buckets. 
Then, the conditional quantile $y_{\tau}=\mu_{\tau}(x)$ is taken based on each respective bucket. 
This allows quantile regression to operate on observed data and construct predictive models.

\subsubsection{Targeted Quantiles}
\label{sec:quantiles}
One strength of quantile regression is its ability to look at different quantiles of the dataset. In this paper, we focus on constructing models for the 95th percentiles of the collected \highlightedit{latency-related} measurements to cover broad aspects of (\NLM, \AFM) performance. (A higher percentile
means that a model explains more of the worst-case data.
Ideally we could present 99th percentile data, but with our current 5-minute sample windows, many buckets do not contain
sufficient samples to compute that percentile.)
% all of which are important aspects of system performance analysis. 
We refer to the model fitted to the 95-th quantile of each network-metric bucket as the
‘95th quantile regression’ model, or QR95 for short.
(\srf{sec:qr-thresholds} shows an example of how different quantiles lead to different models.)

\highlightedit{
For delivery-rate AFMs, since SLOs focus on worst-case behavior, we
collected the p1 rates (i.e, worse than 99\% of the samples) and then
fit models to the 5th percentile of the measured p1 rates, not the 95th.
%Conversely, we choose to inspect the 5th percentile of delivery rate \AFMs, as we also want to focus on the worst-case data, which are the cases where delivery rates are smaller.
}

\subsubsection{Asymmetric Bias}\label{ssec:directional_loss}
During training (model creation), we want
to allow users of our methodology to tune its objective towards overprediction
or underprediction, depending on their understanding of relative risk.
Therefore, instead of traditionally symmetric, least-square-based estimation methods,
we introduce an asymmetric form of the least-squares objective (inspired by the formulation in ~\cite{hydrologic_modeling_loss}).
%\footnote{Our asymmetric objective is inspired by its use in other fields.
%For example, in hydrologic modeling, when estimating runoff-water flows, it is often
%beneficial to overpredict flows during flood seasons, and underpredict flows during drought seasons~\cite{hydrologic_modeling_loss}.  
%\fixthis{check this citation/statement}}
Specifically, we define %an \textit{asymmetric} least-squares objective, 
the \emph{asymmetric mean squared error (AMSE)} as:
%\begin{align*}
%AMSE = 
%       2*\Big(\alpha * \frac{1}{N_o}\sum_{i=1}^{N_o}[max(0, f(x_i)-y_i)]^2 \\ 
%       +(1 - \alpha) * \frac{1}{N_u}\sum_{i=1}^{N_u}[min(0, f(x_i)-y_i)]^2 \Big)
%\numberthis \label{eq:amse}
%\end{align*}
\arxivedit{
\begin{equation}
\small
\begin{split}
    AMSE = 
       2\Big ( \alpha  \frac{1}{N_o}\sum_{i=1}^{N_o}[\max(0, f(x_i)-y_i)]^2  \\
       +(1 - \alpha)  \frac{1}{N_u}\sum_{i=1}^{N_u}[\min(0, f(x_i)-y_i)]^2 \Big)
           \label{eq:amseX}
\end{split}
\end{equation}
}
where $y_{i}$ is the ground truth of the target variable, $f(x_{i})$ is the model prediction, $N_{o}$ is the number of samples the model overpredicted, and $N_{u}$ is the number of samples the model underpredicted.
$\alpha$ is an adjustable weight that allows the regression model to favor overprediction or underprediction during training.
When $\alpha$ is larger 0.5, 
AMSE penalizes overprediction more; where overpredicted samples are captured by the $max$ term, and vice versa.
AMSE falls back to symmetric least-squares when $\alpha=0.5$.

When evaluating a predicted model against test data, we wanted a normalized metric,
so we use
the relative, square-rooted form of AMSE, \emph{relative asymmetric root mean squared error (rARMSE)}.
%, to evaluate the prediction accuracy on unseen test data. 
Taking the root of the squared error makes rARMSE have the same units as the target variable and easier to interpret the magnitude of the prediction errors.
Calculating the error as a ratio (with denominator $y_i$) makes the error scale-independent, allowing it to be applied across different \AFMs without setting specific error thresholds for each individual \AFM.
rARMSE can be written as:
% \begin{equation}
% \small
% \begin{split}
% rARMSE = 
%       \sqrt{2*\alpha * \frac{1}{N_o}\sum_{i=1}^{N_o}[max(0, f(x_i)-y_i)/y_i]^2} \\ 
%       \overline{+(1 - \alpha) * \frac{1}{N_u}\sum_{i=1}^{N_u}[min(0, f(x_i)-y_i)/y_i]^2}
% \label{eq:rarmse}
% \end{split}
% \end{equation}
\arxivedit{
\begin{equation}
\small
\begin{split}
rARMSE = 
       2\Big(\alpha  \frac{1}{N_o}\sum_{i=1}^{N_o}[\max(0, (f(x_i)-y_i)/y_i)]^2\\+(1 - \alpha)  \frac{1}{N_u}\sum_{i=1}^{N_u}[\min(0, (f(x_i)-y_i)/y_i)]^2\Big)^{\frac{1}{2}}
\label{eq:rarmse}
\end{split}
\end{equation}
}

\subsubsection{Linear and Non-linear quantile regression}~\label{sssec:feature_transformation}
We choose to fit (\NLM, \AFM) relationships on both a linear model and a queueing (M/D/1) model to reflect the lightly-loaded and moderate-queuing regimes of a network (Figure~\ref{fig:regimes-hypothetical}).
We perform quantile regression twice, once on each model, and choose the model with the lowest AMSE. 

% We currently assume that we can potentailly model the entire network as a single M/D/1 queue.

% That assumption depends on there being only one significant source of queueing on
% each network path, which is often (but not always) the case in moderate-queueing
% regimes.  

%On the other hand, it is also equally useful to include the theoretically supported models. To this end we included the queueing theory model, the M/D/1 queue in particular, which suggests that service delay and utilization follow a reciprocal relationship. 
% In an M/D/1 queue, the system is assumed to have a single server, the arrival rates are determined by a random process, and the service time of each request is fixed. 
%The relationship between service delay and queue utilization of an M/D/1 queue can be translated into the relationship between transfer latency and link utilization, where we model the network as one giant queue. 

For an M/D/1 queue, service delay (such as transmit latency)
and utilization follow a reciprocal relationship~\cite{qt}. %\citeneeded{}
When fitting our data to an M/D/1 model, we first perform feature transformation based on Equation~\ref{eq:feature_transformation}. Feature transformation allows us to treat non-linear models as if they are using linear features, allowing us to solve as linear regresssions, which is more efficient.
\arxivedit{
Note that linearity is considered with respect to the original features or with respect to the transformed features, whereas the queueing theory-based model assumes a non-linear relationship between utilization and delay.
}
\begin{equation}\label{eq:feature_transformation}
    Percentile_{\tau}(y) = \beta\frac{x}{1-x} + c = \beta Q + c
\end{equation}

\subsection{Combined algorithms}
\label{sec:combined}
To combine Kneedle and quantile regression,
for each (\NLM, \AFM) pair of interest, we derive a predictive model by:
\begin{enumerate}
    \item Applying the Kneedle algorithm to detect and record a knee, if any.
    \item If there is a knee, creating a subset of the data points for which the \NLM
    is smaller than one bucket before the knee, to ensure that there are no false negatives due to bucketization. If there is no knee, use the entire dataset as this
    subset.
    (The offset of one bucket is arbitrary, but appears to work.)

    \item For the subset of (\NLM, \AFM) pairs selected in the previous step, we perform
    quantile regression twice, using both linear and queueing-theory models.
    \item We report the model (linear or queueing theory) with the lowest rARMSE. If no models have a confidence score above a threshold (\srf{sec:error-threshold}), we assume that the (\NLM, \AFM) pair has no clear relationship. 
\end{enumerate}

\parab{Understanding model stability}:
The best model for a given fabric might vary over time.
To understand this variability, we can derive a model (for a given
\NLM-\AFM pair) using a subset of the input data covering time spans of
various durations, and then validating these models against held-back data from
other time spans, i.e. splitting the input data into \emph{training and testing sets}.
This helps network designers and operators understand how frequently they
would need to update their predictive models, to avoid stale results.

\arxivedit{
Our methodology combines a data-driven approach that utilizes training and testing sets, with an analytical framework based on queueing theory models. The key insight to this combined learning-based + analytical model approach is that we can adapt certain predictive models (e.g. the M/D/1 queueing model) that are informed by domain-specific knowledge of network behavior, instead of using generic regression techniques or deep neural networks. 
On the other hand, the integration of data-driven techniques then ensures that our predictive models remain robust to changes in \NLM/\AFM patterns, remaining practical and actionable for network operators in real-world scenarios. 
}

\section{Production Fabric Case Study}
\label{sec:eval_case_study}
Before presenting a broad analysis across 19 production fabrics, code-named $A..S$, in
\srf{sec:multi-fabric-eval}, we present
an in-depth analysis of fabric \emph{C}, which contains 51 active aggregation blocks.
We chose \emph{C} because its operators have chosen to allow several aggregation blocks to run at persistent
high utilizations, and so this fabric illustrates several interesting aspects of our methodology.

For fabric \emph{C} we looked at how well we can build predictive models (via training and testing), 
whether the choice of model (linear or queueing-based) is stable from week to week,
and which \NLMs are the best predictors.
We present results for both low-QoS and medium-QoS flows; the fabric preserves sufficient bandwidth
for high-QoS flows that these results are uninteresting (\ie they always perform well).
We looked at both \NLMs representing inter-aggregation-block links and aggregated adjacencies
(\srf{sec:inter-block-predictions}) and per-block
\NLMs representing intra-block links (\srf{sec:intra-block-eval}).

 We aim to answer the following questions:\\
 \textbf{Q1:} Is it feasible to create fabric-wide predictions? (\ie using \NLMs that represent fabric-wide network performance) (\srf{sec:inter-block-predictions})\\
 \textbf{Q2:} Which \NLMs are good predictors, and are the same \NLMs good predictors in all cases? 
 (\srf{sec:best-nlms})\\
 \textbf{Q3:} Is the best choice of models stable over periods of weeks or longer? (\srf{sec:stability-eval})\\
 \textbf{Q4:} In which circumstances are linear or queuing-based models better predictors? (\srf{sec:stability-eval}) \\

 We chose to analyze these \NLMs:
 \begin{itemize}
    \item Link Utilization: utilization, measured over one sample period (5 minutes) of a given link.
    \item Maximum Link Utilization (\emph{MLU}): maximum, during one sample period, of all link utilizations in the fabric.
    \item Average Link Utilization (\emph{ALU}): average, during one period, of all link utilizations in the fabric.
    \item Adjacency Utilization: utilization aggregated over all links between a pair of blocks.
    \item Maximum Adjacency Utilization (\emph{MAU}): maximum, during one sample period, of all adjacency utilizations in the fabric.
    \item Average Adjacency Utilization (\emph{AAU}): average, during one period, of all adjacency utilizations in the fabric.
    \item \emph{Jain}: Jain's fairness index~\cite{jain1984quantitative}, computed over the set of link utilizations, as a proxy
    for ``load imbalance'' across those links.
    \item \emph{P5-P95-distance}: difference between the 5th and 95th percentile link utilizations
    (an alternate measure of imbalance).
 \end{itemize}
These metrics are derived from the metrics and metadata summarized in \trf{table:netdata}.
 
\subsection{Model-training and testing parameters}
~\label{ssec:setup}
Our dataset covers two months.  For most of these results, we trained models on 4 weeks of data, and
tested them on data from the following 2 weeks.  
% (The week-to-week stability results in \srf{sec:stability-eval} used 1-week testing windows.)
We split the \NLM data into 20 buckets (5\%/bucket) for maximum utilizations, and into 100 buckets (1\%/bucket)
for mean utilizations, since the latter datasets had smaller domains, and we needed smaller buckets to have enough
that contained sufficient samples.
For knee detection, we set the curvature threshold $\mathcal{C}=0.5$.
We declared a model to be ``accurate'' during testing if the $rARMSE < 0.15$.
\srf{sec:error-threshold} discusses the choice of this threshold.

Throughout this section, we present QR95 results, as defined in \srf{sec:quantiles}, unless otherwise
noted.

\begin{figure}
 \begin{minipage}{0.5\textwidth}
     \centering
     \includegraphics[width=\linewidth]{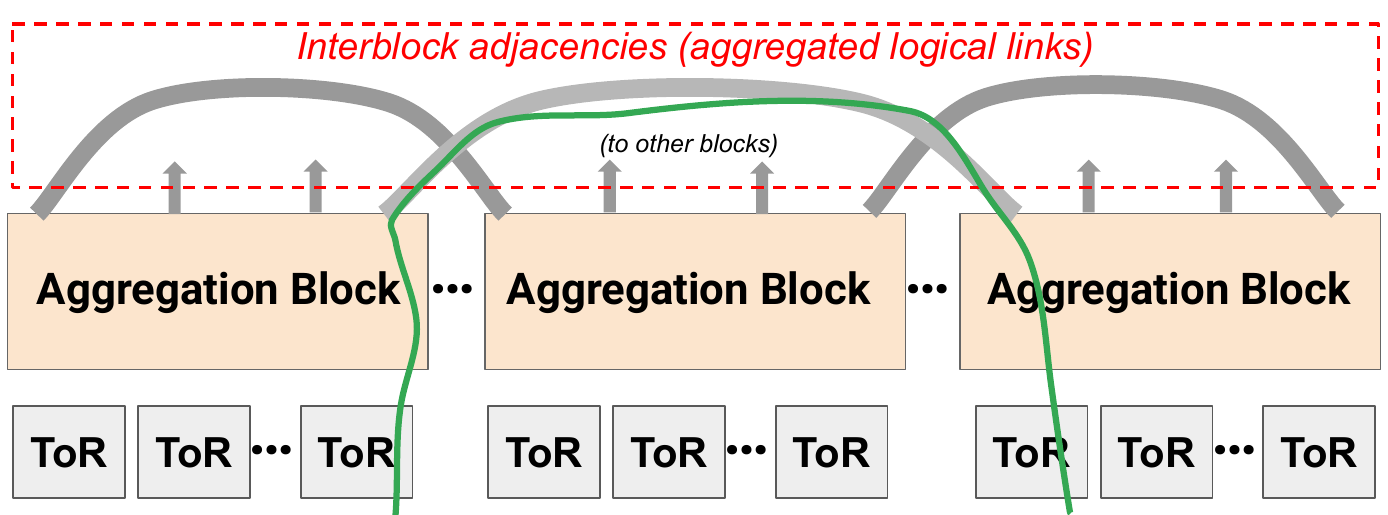}
     \\ \small{Green line shows path of an inter-block RPC}
     \caption{Links used for inter-block \NLMs}
     \label{fig:interblock-adjacencies}
 \end{minipage}
 \begin{minipage}{0.5\textwidth}
     \centering
         \includegraphics[width=.95\textwidth]{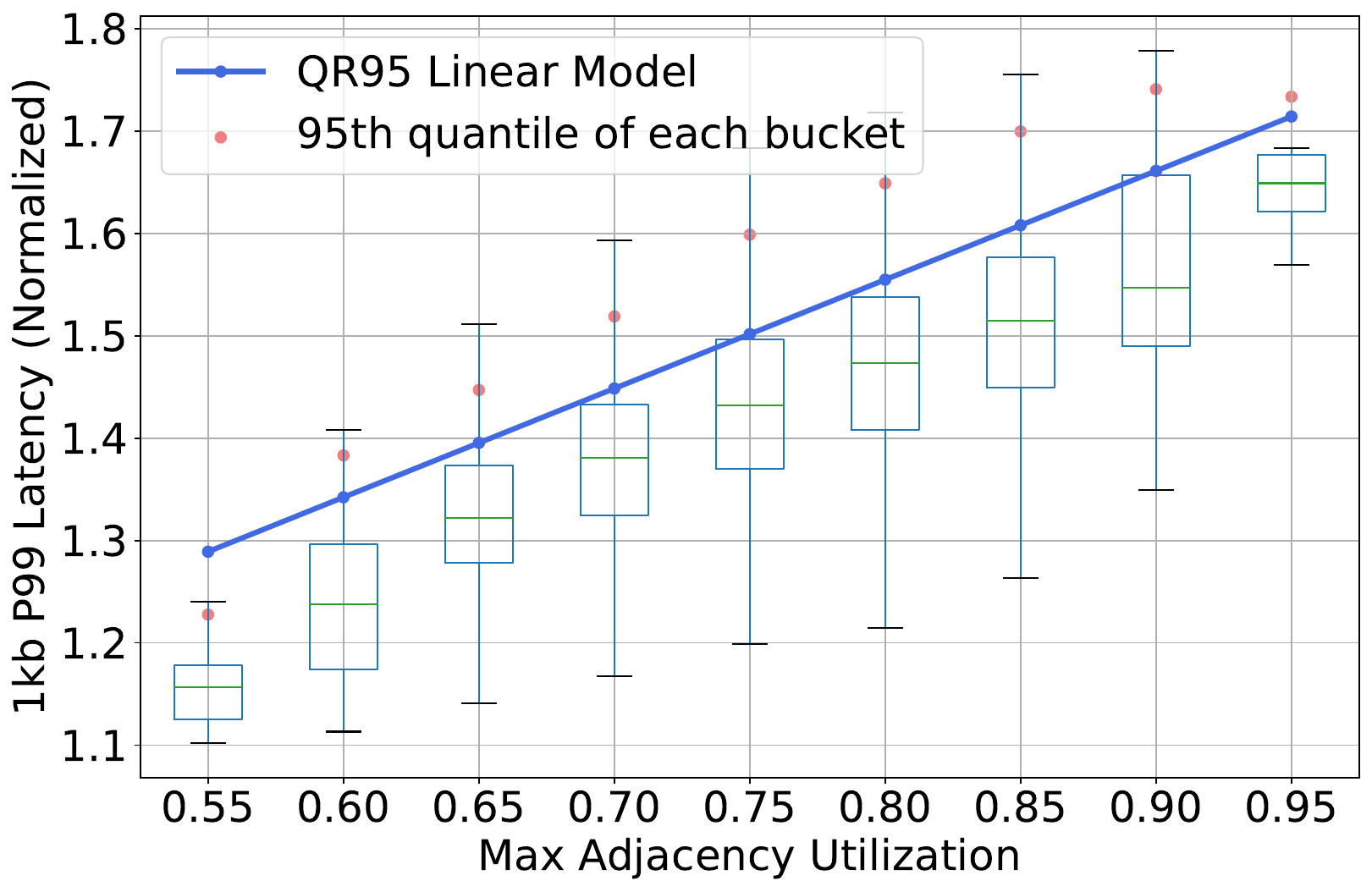}\\
     \caption{Fabric \emph{C}, 64kb Tail Transmit Latency, medium QoS, \NLM$=$ MAU.}
     \label{fig:case_study_med_medqos}
 \end{minipage}
\end{figure}

\subsection{Fabric-wide (inter-block) predictions}
\label{sec:inter-block-predictions}

In this section, we look at \AFMs for RPCs that cross between aggregation blocks, using \NLMs only from the
aggregated links (``adjacencies'') between blocks, as illustrated in \frf{fig:interblock-adjacencies}.

\begin{figure*}[ht!]
     \centering
     \begin{subfigure}[]{0.5\textwidth}
         \centering
         \includegraphics[width=.95\textwidth]{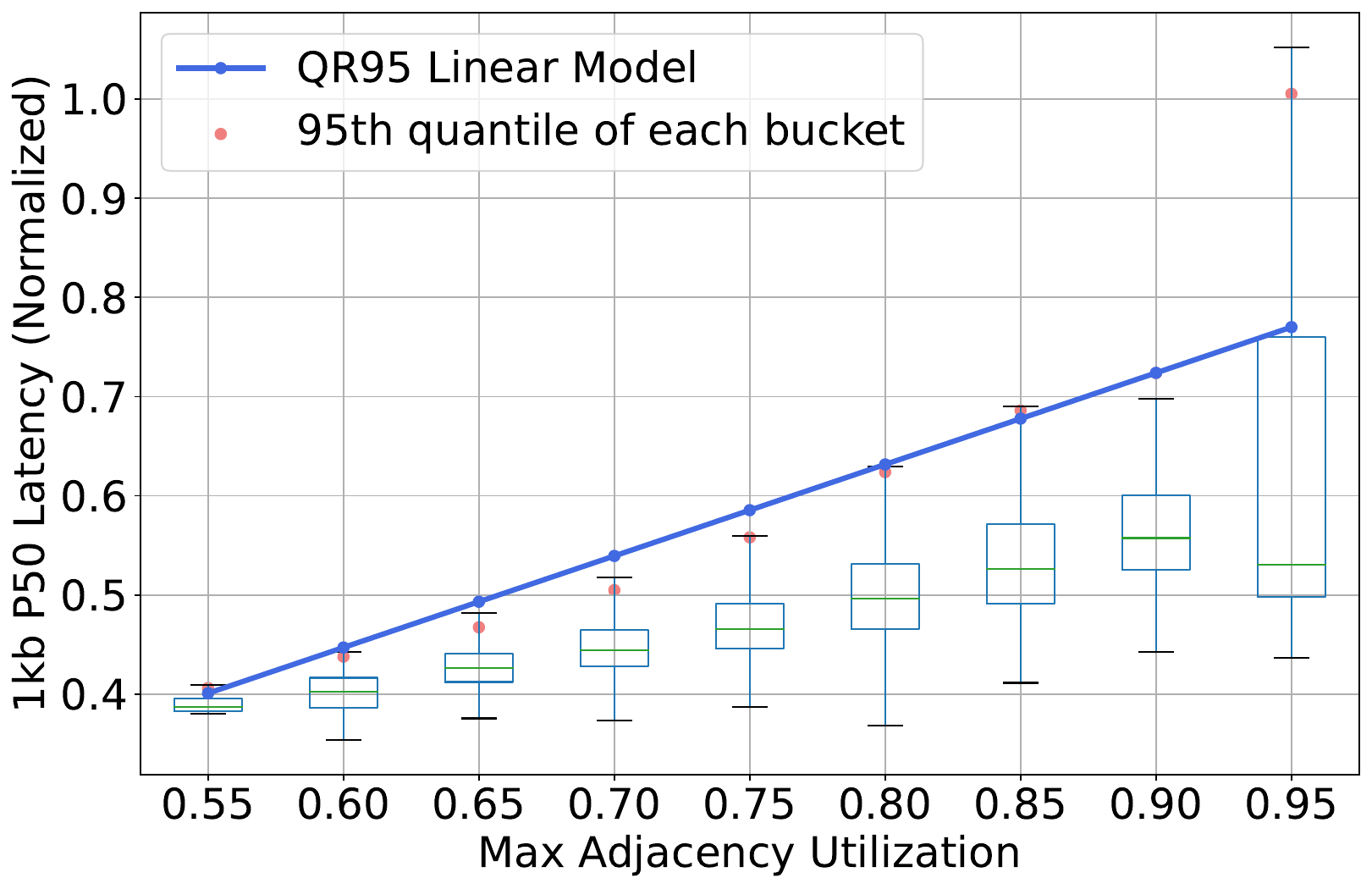}
         \caption{1kb Median Transmit Latency.}
     \end{subfigure}%
     \begin{subfigure}[]{0.5\textwidth}
         \centering
         \includegraphics[width=.95\textwidth]{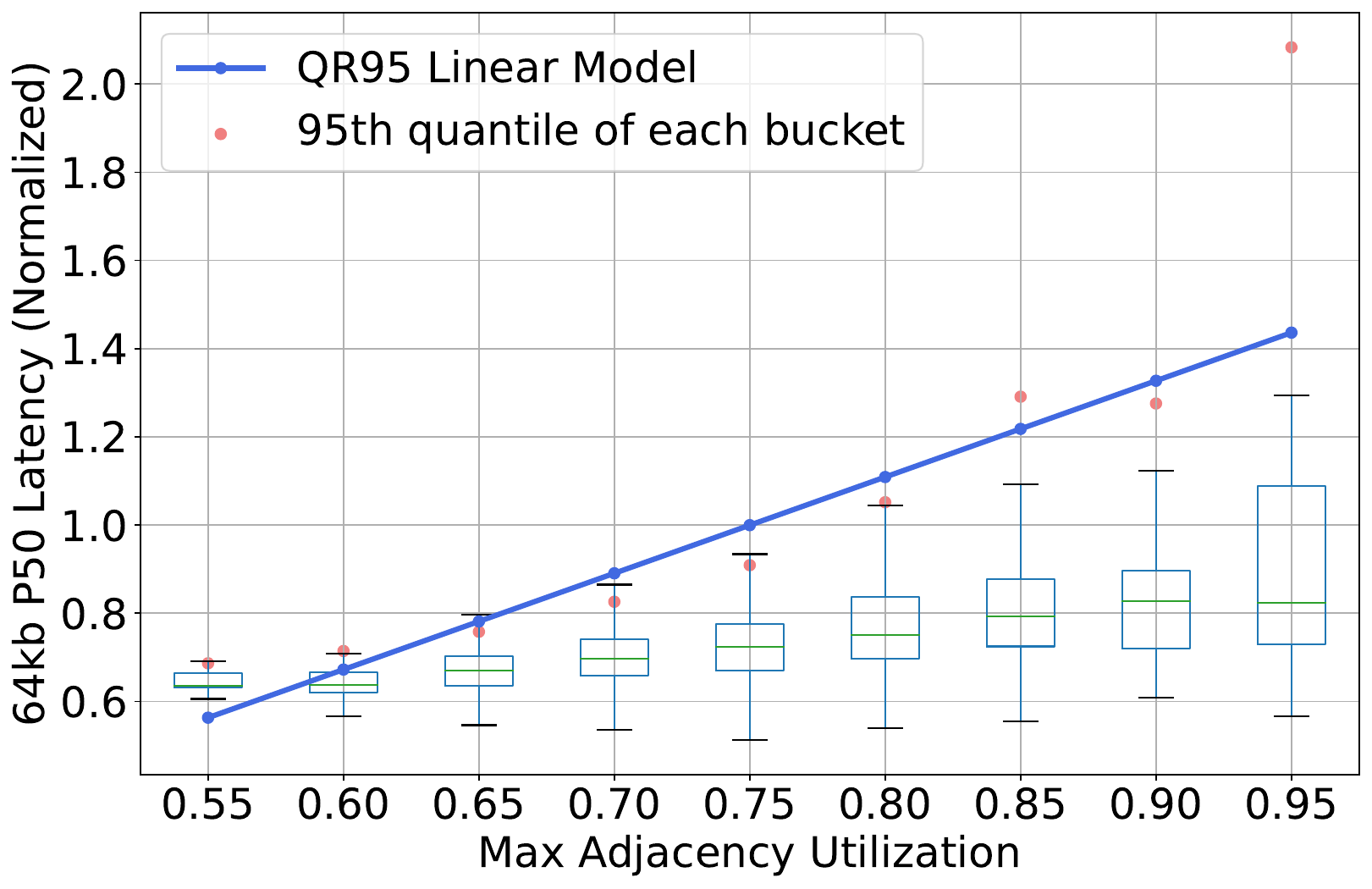}
         \caption{64kb Median Transmit Latency.}
     \end{subfigure}
     \caption{Fabric \emph{C} models: p50 \AFMs, low QoS, \NLM$=$ MAU.}
        \label{fig:case_study_median_biweekly}
\end{figure*}
\begin{figure*}[ht!]
     \centering
     \begin{subfigure}[]{0.5\textwidth}
         \centering
         \includegraphics[width=.95\textwidth]{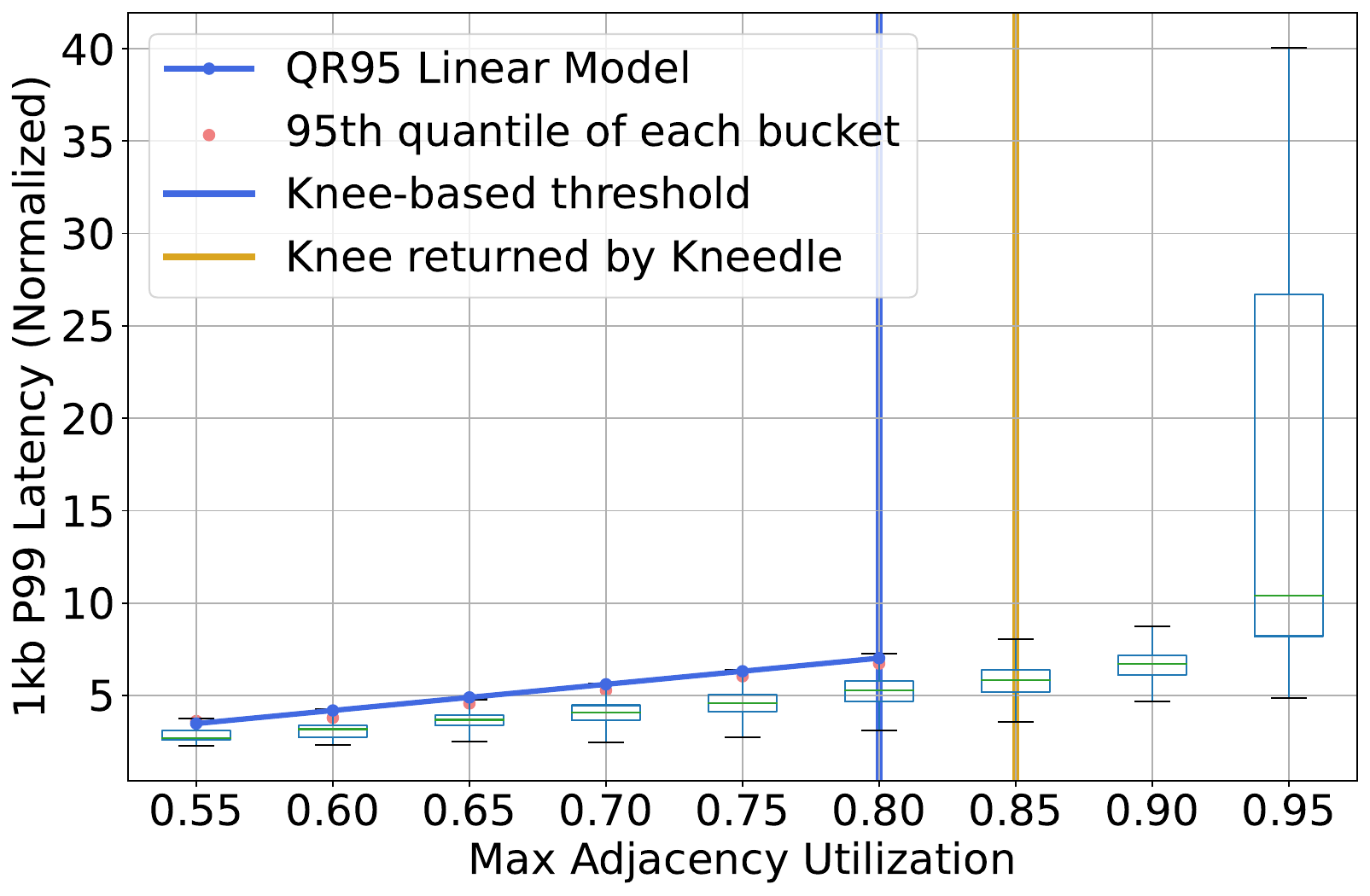}
         \caption{1kb Tail Transmit Latency.}
         \label{sfig:1kbp99lowboxplot}
     \end{subfigure}%
     \begin{subfigure}[]{0.5\textwidth}
         \centering
         \includegraphics[width=.95\textwidth]{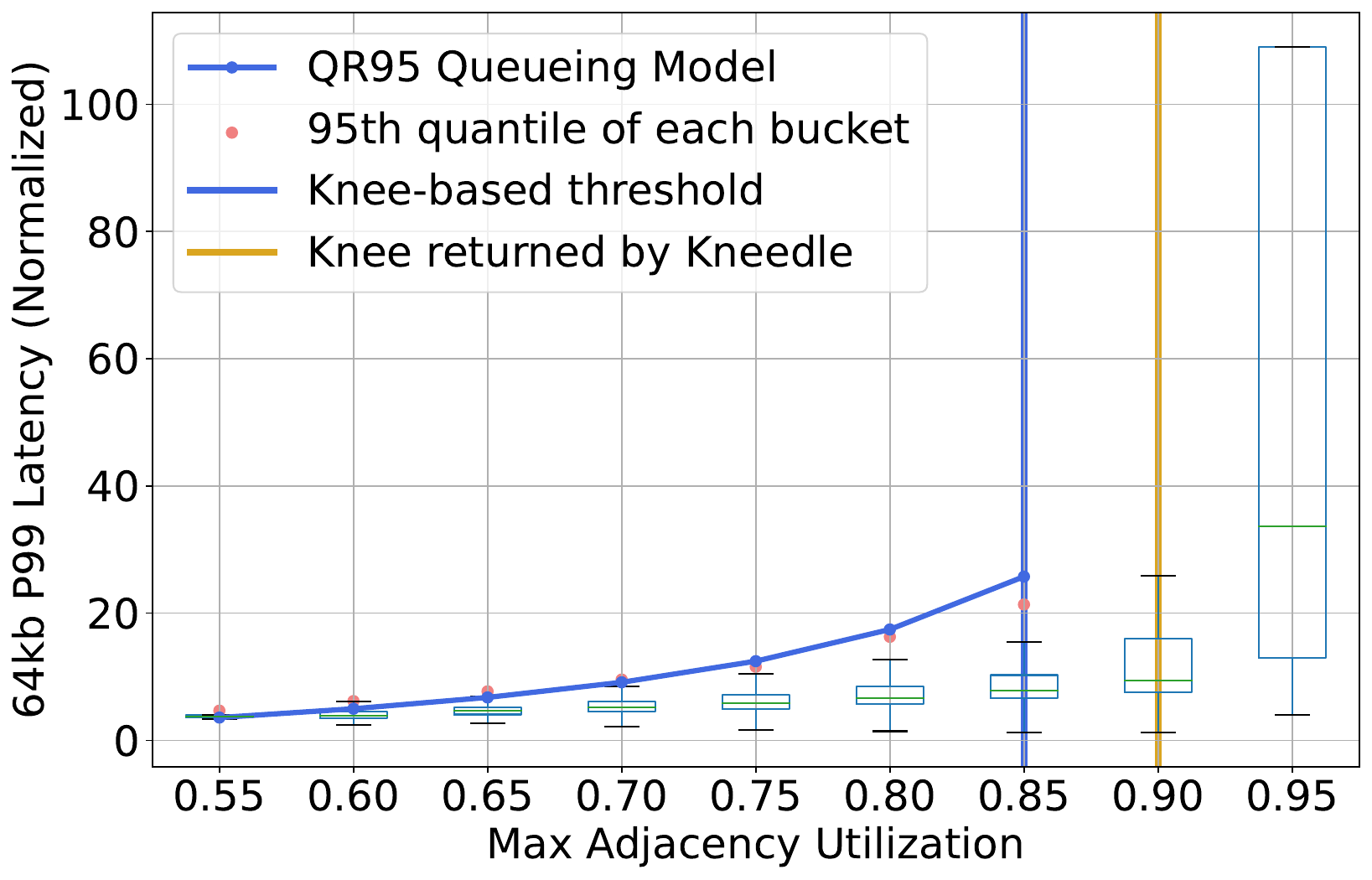}
         \caption{64kb Tail Transmit Latency.}
     \end{subfigure}
          \caption{Fabric \emph{C} models: p99 \AFMs, low QoS, \NLM$=$ MAU.}

        \label{fig:case_study_tail_biweekly}
\end{figure*}
Figures~\ref{fig:case_study_median_biweekly} and~\ref{fig:case_study_tail_biweekly} 
present detailed results for low-QoS flows (the QoS most affected by network utilization).
\textbf{These and all similar figures are normalized to consistent but arbitrary baselines.}
Both show results using MAU as the \NLM predictor
(\srf{sec:best-nlms} discusses other \NLMs).
These graphs show box plots of the underlying
test data, vertical lines for the upper limit of the regression subset (see \srf{sec:combined}), and curves for
the linear or queueing-based models we predicted.

\frf{fig:case_study_median_biweekly} shows that for p50 (median) \AFMs, we obtain good fits with linear models.
\frf{fig:case_study_tail_biweekly} shows that for p99 (tail) \AFMs, a queueing-based model is a much
better fit, and that \AFMs increase rapidly after the knees.

We did not detect any knees for medium-QoS flows, likely because the network sufficiently protects these flows.
\highlightedit{
\frf{fig:case_study_med_medqos} illustrates this for median-QoS RPC tail latency.
}

Note that we tried training models on data grouped by congestion-control algorithm (CCA), since different flows in our
dataset use different CCAs, but this did not lead to accurate  models. 
\highlightedit{For various other \AFMs,
such as the number of timeouts, we could not construct high-confidence models.
In a well-managed network, these events are rare, even at high utilization.}

\subsubsection{Best \NLMs}
\label{sec:best-nlms}
\highlightedit{
For Fabric~\emph{C}, different \NLMs provide the highest prediction
accuracy (lowest error) in different cases.
Maximum adjacency utilization (MAU)
was consistently the best predictor for low-QoS traffic, 
but generally had poor accuracy for high-QoS traffic (possibly because, for high QoS, end-host delays rather than network delays, dominate~\cite{AgarwalEtAL2022}).
Medium-QoS \AFMs were best predicted by MAU, 
except that P95\_P5\_dist was a slightly better predictor, than MAU, for
64kb-RPC median latencies.
}

In \srf{sec:fabricwide}, we plot accuracies across multiple fabrics, and show
that for a given \AFM, the \NLM that yields the lowest prediction error changes fabric-to-fabric. 

\subsubsection{Stability}
\label{sec:stability-eval}
\noindent \textbf{\\Prediction stability. }
Ideally, models would be stable over time -- \ie a model that worked during the most recent two weeks (linear or queueing) would
still work for the next two weeks.  
To evaluate stability, we applied a sliding window, where we trained a new model starting every two weeks, using four weeks of training data, and tested it on the fifth and sixth week's data.
We used an additional month of Fabric \emph{C} data to obtain sufficient results for four consecutive windows.

\begin{figure*}[ht!]
     \centering
     \footnotesize
     \begin{subfigure}[t]{0.24\textwidth}
         \centering
         \includegraphics[width=\textwidth]{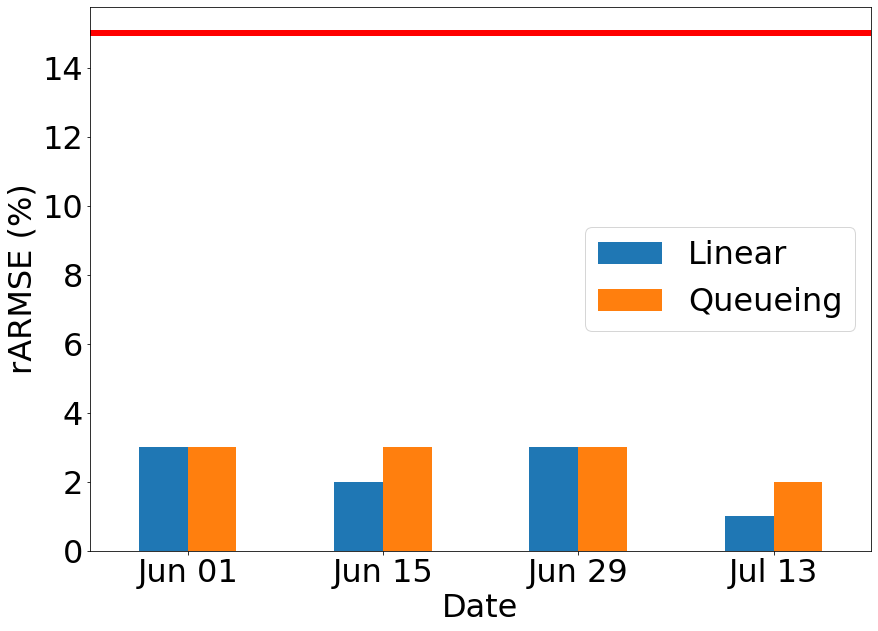}
         \caption{1kb RPC Latency, P50.}
         \label{fig:stability-1kbp50-medium}
     \end{subfigure}%
     \begin{subfigure}[t]{0.24\textwidth}
         \centering
         \includegraphics[width=\textwidth]{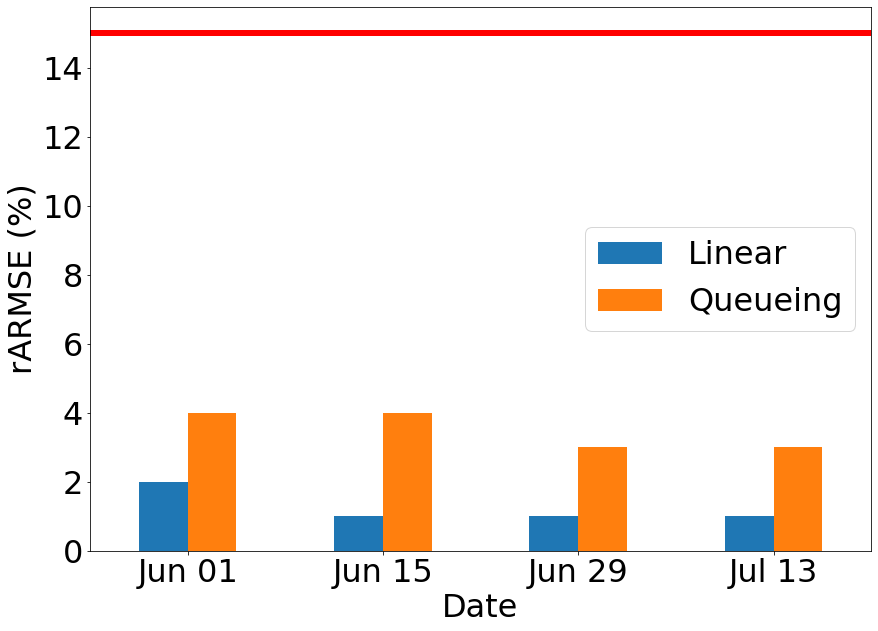}
         \caption{64kb RPC Latency, P50.}
         \label{fig:stability-64kbp50-medium}
     \end{subfigure}
     \begin{subfigure}[t]{0.24\textwidth}
         \centering
         \includegraphics[width=\textwidth]{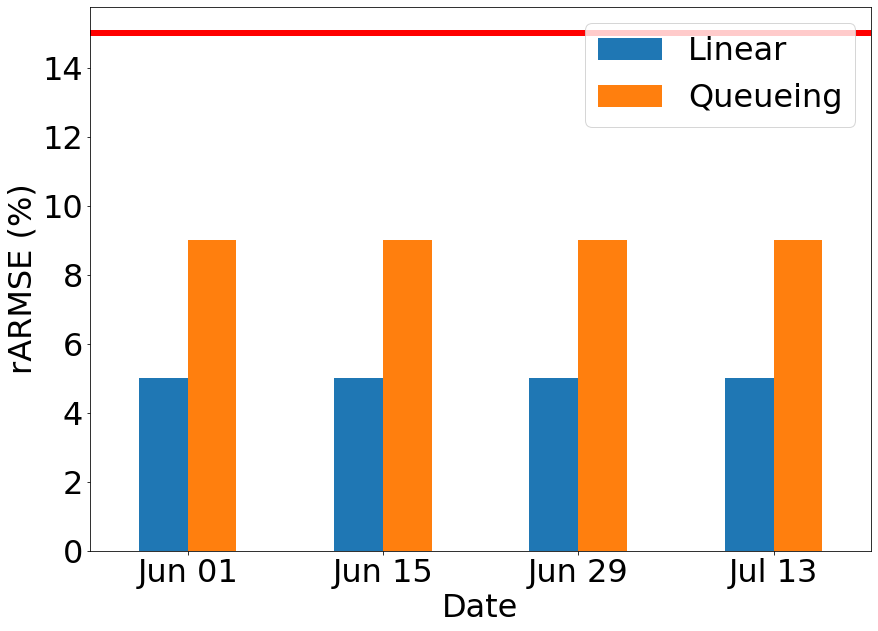}
         \caption{1kb RPC Latency, P99.}
         \label{fig:stability-1kbp99-medium}
     \end{subfigure}%
     \begin{subfigure}[t]{0.24\textwidth}
         \centering
         \includegraphics[width=\textwidth]{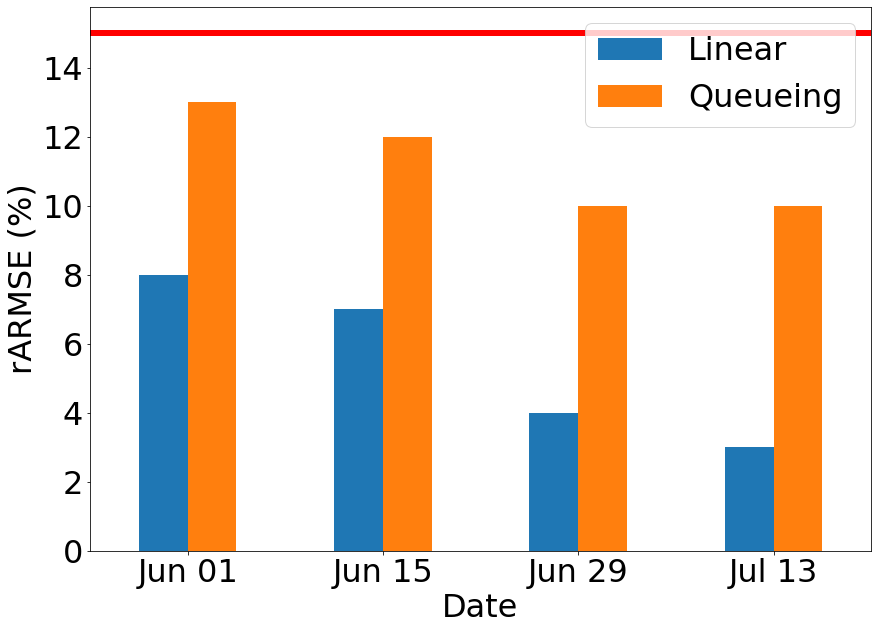}
         \caption{64kb RPC Latency, P99.}
         \label{fig:stability-64kbp99-medium}
     \end{subfigure}
        \\ Red line indicates the rARMSE threshold ($15\%$).
        \NLM $=$ MAU in these graphs
        \caption{Week-to-week stability, medium-QoS flows.}
        \label{fig:stability-medqos}
\end{figure*}

\frf{fig:stability-medqos} shows the corresponding results for medium-QoS flows.  These predictions
are more stable and generally more accurate than those in \frf{fig:stability-lowqos} -- always well within
our 15\% threshold -- and the linear model is consistently more accurate than the queueing-based model,
presumably because the medium-QoS flows are less likely to experience queueing.

\begin{figure*}[ht!]
     \centering
     \footnotesize
     \begin{subfigure}[t]{0.24\textwidth}
         \centering
         \includegraphics[width=\textwidth]{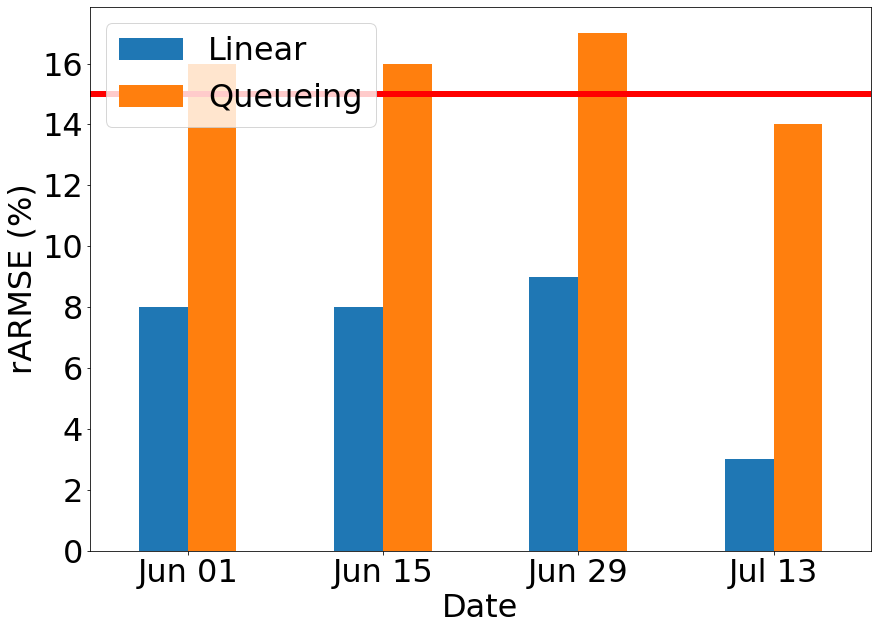}
         \caption{1kb RPC Latency, P50.}
         \label{fig:stability-1kbp50}
     \end{subfigure}%
     \begin{subfigure}[t]{0.24\textwidth}
         \centering
         \includegraphics[width=\textwidth]{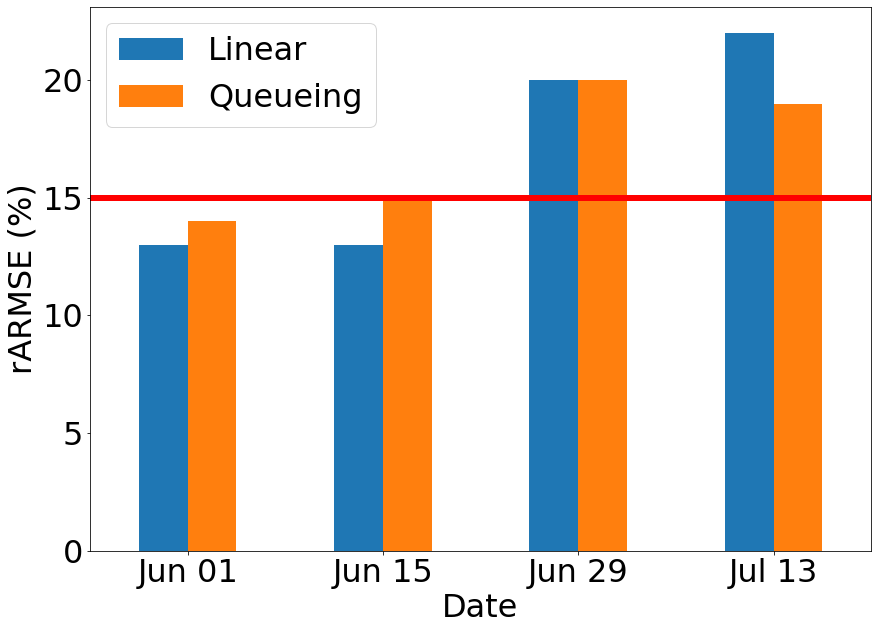}
         \caption{64kb RPC Latency, P50.}
         \label{fig:stability-64kbp50}
     \end{subfigure}
     \begin{subfigure}[t]{0.24\textwidth}
         \centering
         \includegraphics[width=\textwidth]{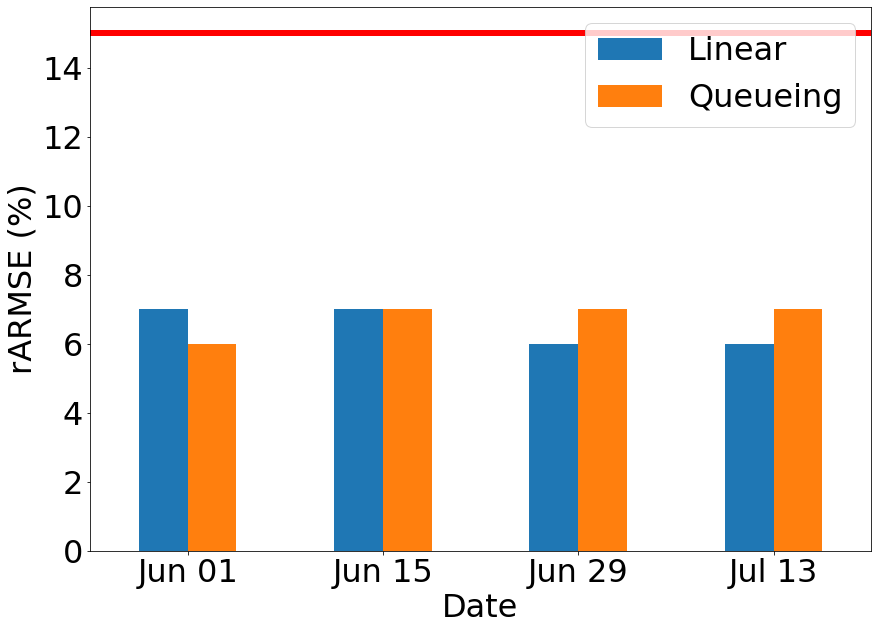}
         \caption{1kb RPC Latency, P99.}
         \label{fig:stability-1kbp99}
     \end{subfigure}%
     \begin{subfigure}[t]{0.24\textwidth}
         \centering
         \includegraphics[width=\textwidth]{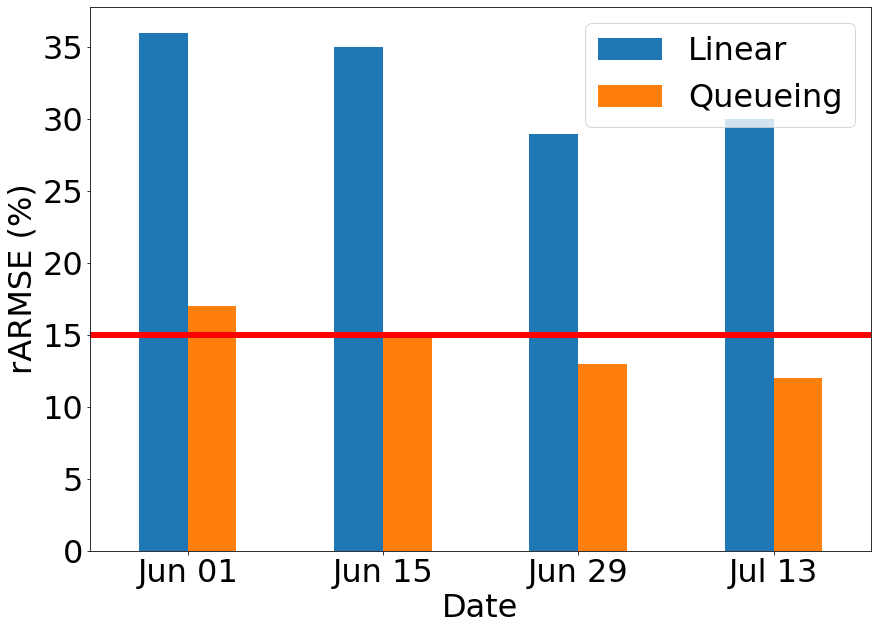}
         \caption{64kb RPC Latency, P99.}
         \label{fig:stability-64kbp99}
     \end{subfigure}
        \\ Red line indicates the rARMSE threshold ($15\%$).
        \NLM $=$ MAU in these graphs
        \caption{Week-to-week stability, low-QoS flows.}
        \label{fig:stability-lowqos}
\end{figure*}

\frf{fig:stability-lowqos} shows that, for low-QoS flows, our algorithm consistently chooses a linear model for small-RPC p50 latency (\frf{fig:stability-1kbp50}). It consistently chooses a queueing model for large-RPC p99 latency (\frf{fig:stability-64kbp99}), presumably because large RPCs take more round trips to complete, so DCN conditions
are more likely to trigger congestion-avoidance mechanisms.
For large-RPC p50 (\frf{fig:stability-64kbp50}) and small-RPC p99 (\frf{fig:stability-1kbp99}) \AFMs, neither
model consistently ``wins.''   However, in most cases, we can find a model that meets our 15\% error 
threshold.\footnote{We currently cannot explain why the p50 models are less likely than p99 models to be within
the error threshold, for low-QoS flows, even though for medium-QoS flows we do see higher errors for p99.
Note that medium-QoS flows are likely to be from different applications, rather than being higher-priority
versions of low-QoS flows, so some differences are inherent.}

\noindent \textbf{\\Knee stability. }
\highlightedit{
In cases where we detect a knee, is the knee stable?
%For the low-QoS \NLM-\AFM relationships where we detected a knee, we would also like the position of the knee to remain stable.
%Similarly, we
We applied a one-week sliding window, across 6 weeks, to see if
knees appear, disappear, or change position between weeks.
%there are any value changes, appearances, or disappearances of knees for adjacent time windows.
}

\highlightedit{
Indeed, in most cases for this fabric (for both small and large RPC latencies)
for P99 \AFMs for low-QoS flows, the knees we detected were stable
at MAU=90\%.   In one case (July 6, small RPC latencies) the knee moved to
MAU=95\%; this could be a bucket-quantization effect.
For delivery rates, we found a knee only for the week of July 6.
}

\subsection{Block-level Predictions}
\label{sec:intra-block-eval}

In this section, we look at \AFMs for RPCs that stay within aggregation blocks, using \NLMs only from the individual
(non-aggregated)
internal links within each block, as illustrated in \frf{fig:intrablock-links}.
Note that we do not include \NLMs from ToR uplinks or downlinks.

\begin{figure}[ht!]
    \begin{minipage}{.49\textwidth}
     \centering
     \includegraphics[width=\linewidth]{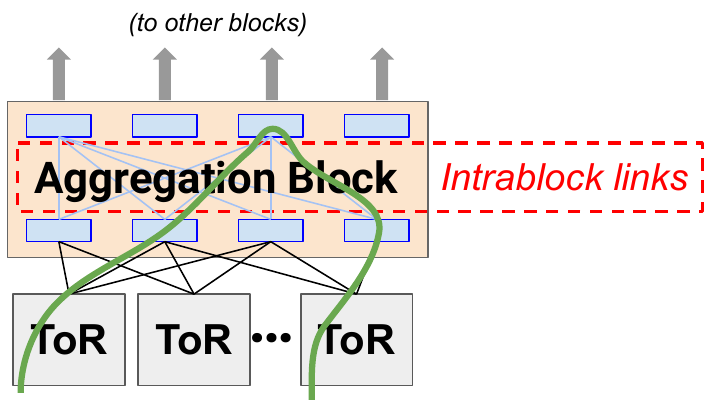}
          \\ \textit{\small{Green line shows path of an intra-block RPC.}}
     \caption{Links used for intra-block \NLMs}
     \label{fig:intrablock-links}
    \end{minipage}
    \begin{minipage}{.49\textwidth}
         {\centering
         \small
         \includegraphics[width=\textwidth]{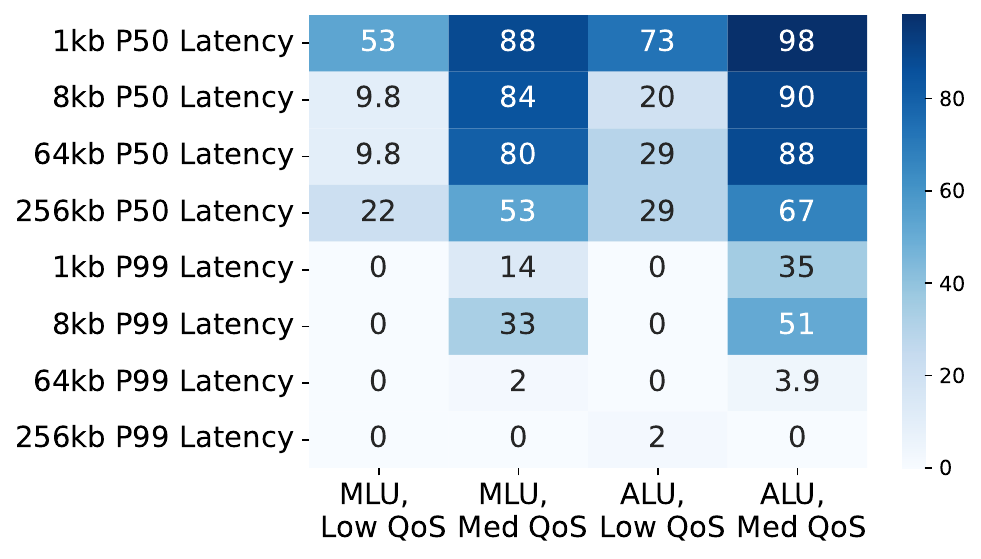}
         }
         \\
         \textit{\small{Cells $=$ \% of all aggregation blocks with models with $rARMSE \le 0.15$.}}
        \caption{Frequency of agg. blocks w/ accurate models.}
        \label{fig:heatmap}
    \end{minipage}

\end{figure}

Figure~\ref{fig:heatmap} shows, for each \AFM,
the percentage of aggregation blocks, in fabric \emph{C}, for which we obtained
high-accuracy models.
The figure shows results for two \NLMs, average link utilization and maximum link utilization; and for both low-QoS and medium-QoS flows.

These results show that, for flows within an aggregation block, medium-QoS latency is far more predictable
than low-QoS latency;
small-RPC latency is usually easier to predict
than large-RPC latency; and median latency is easier to predict than tail latency.   They also show that ALU
is always (for this dataset) a better predictor than MLU.

\section{Multi-fabric analysis}
\label{sec:multi-fabric-eval}
In this section, we present results across all 19 production fabrics in our dataset, to understand
if the results in \srf{sec:eval_case_study} generalize.  Specifically,

\begin{itemize}
\item What are the \NLMs with the highest prediction accuracy? Is it consistent across fabrics?
    (\srf{sec:fabricwide})
   \item Can we accurately predict \AFMs over all fabrics or some of them? (\srf{sec:fabricwide})
    \item Are linear or queueing models consistently better across all fabrics? (\srf{sec:linear-or-queueing})
    \item What fraction of the data is covered within the knee-based thresholds? (\srf{sec:coverage})
\end{itemize}

\srf{sec:per-fabric-block-level} extends the block-level analysis of \srf{sec:intra-block-eval} to multiple fabrics.

\subsection{Fabric-wide (inter-block) predictions}
\label{sec:fabricwide}

We focus here on tail (p99) latencies for small (1kb) and large (64kb) RPCs, as the most interesting metrics for many
use cases. 
Again, all results are for QR95, as defined in \srf{sec:quantiles}.

\begin{figure*}[htb!]
     \begin{subfigure}[]{\textwidth}
          \centering
         \includegraphics[width=\textwidth]{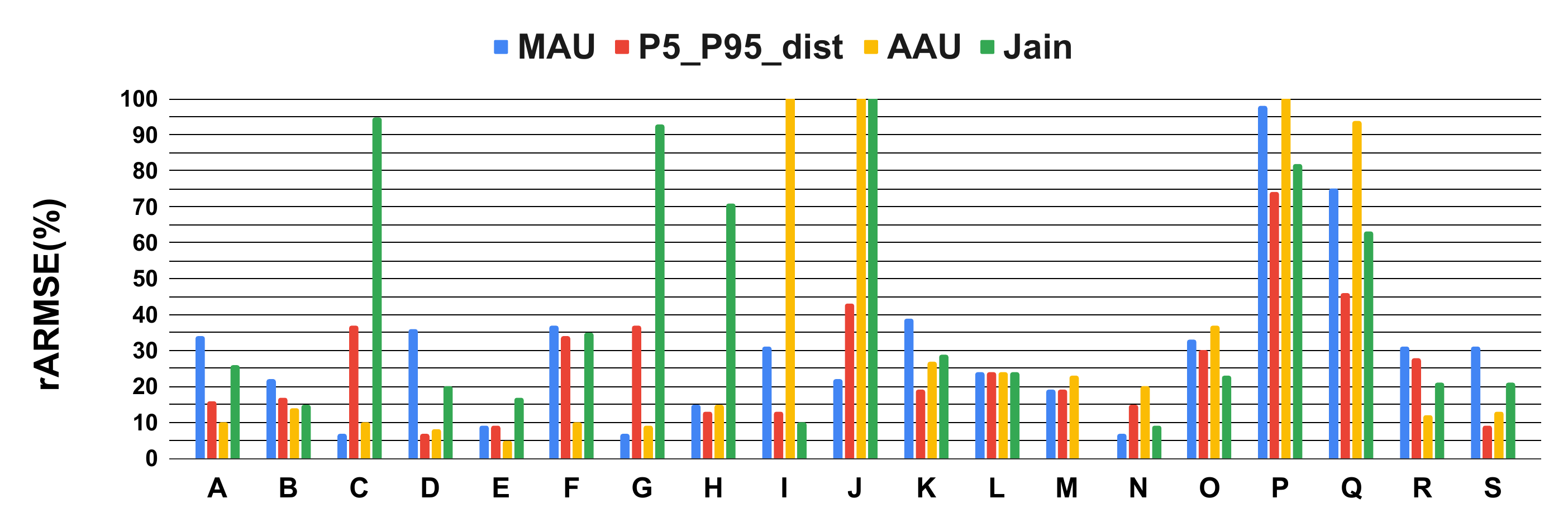}
         \caption{Small (1kb) RPCs.  Some error bars exceed 100\%.}
                 \label{fig:multi-qr95-error-tail-low-small}
     \end{subfigure}
     \begin{subfigure}[]{\textwidth}
          \centering
         \includegraphics[width=\textwidth]{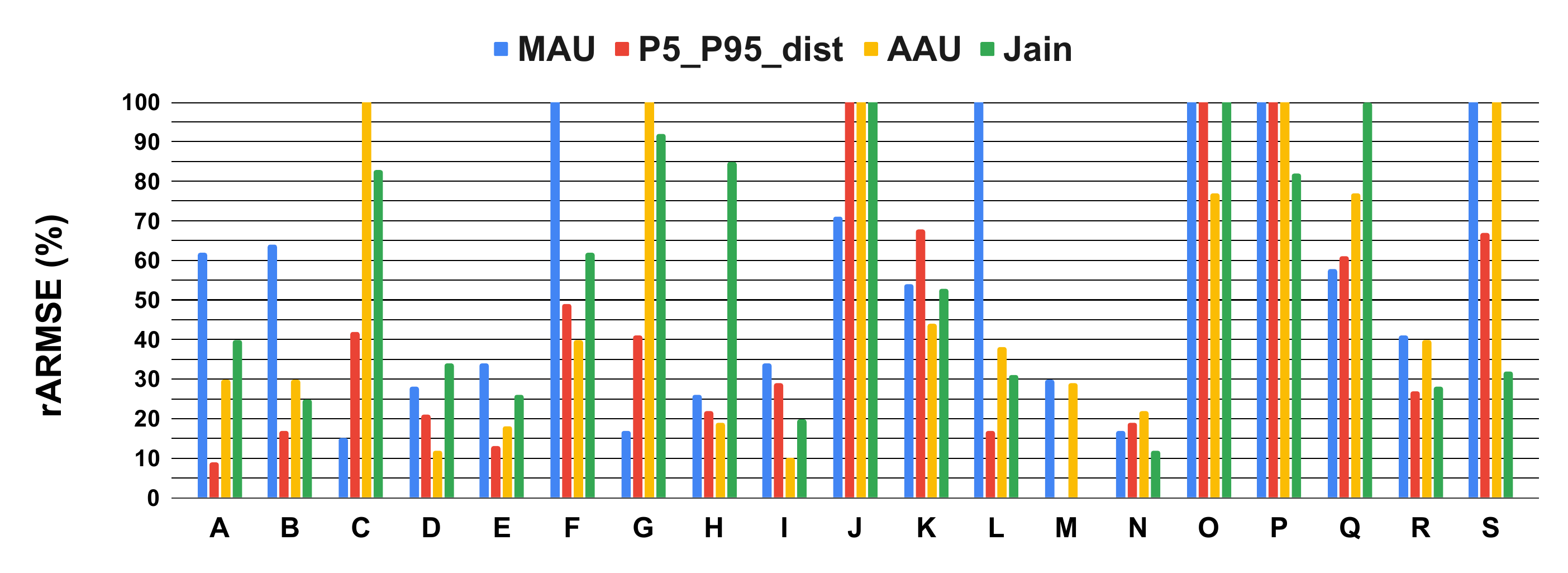}
         \caption{Large (64kb) RPCs.  Some error bars exceed 100\%.}
     \end{subfigure}
          \caption{\small Prediction error for various \AFMs, tail (p99) RPC transmit latency, low QoS.}
        \label{fig:multi-qr95-error-tail-low}
\end{figure*}
 
\frf{fig:multi-qr95-error-tail-low} shows the prediction error (rARMSE) for low-QoS RPC tail transmit latencies, using the model
(linear or queueing-based) that provides the least error, across all 19 fabrics.\footnote{A few bars
are missing because there were insufficient data points to compute a regression, because the knee
found by Kneedle was too small.}
Using the same $rARMSE \le 0.15$ accuracy threshold as in described in \srf{ssec:setup}, we could not obtain
``accurate'' prediction models for some fabrics, regardless of which \NLM we tried.

One takeaway is that while previous work has often used maximum adjacency utilization as an optimization
metric, our results show that this is not always the best choice of \NLM.  
The results also show that the 
\emph{P5-P95-distance} metric for imbalance sometimes works well, while using Jain's index often
works quite poorly.

Overall, no single \NLM reliably predicts low-QoS RPC tail latencies across all of these fabrics, so training models per-fabric seems
essential.  In particular, optimizers for systems such as traffic engineering should have fabric-specific
goals, rather than relying on a fixed metric.

On the other hand, as shown in~\frf{fig:multi-qr95-medqos},
for medium-QoS flows we are much more likely to find accurate models (e.g., for $rARMSE \le 0.15$) than for low-QoS flows.
\srf{sec:more-inter-block} additionally shows that our approach is much more successful at predicting medium-QoS latencies, and median (p50) latencies for low-QoS flows.

\highlightedit{
\frf{fig:multi-qr95-delivery-rate} shows that
%the prediction error of 
for tail (p1) delivery rates for medium-QoS traffic across all 19 fabrics,
we could predict these rates within our 15\% threshold -- except for
Fabrics~\emph{I},~\emph{K}, and~\emph{R}.
(We were unable to extract models for low-QoS delivery rates, because our data included
zero-valued p1 (worst-case) samples, which led to division-by-zero errors
when calculating rARMSE ($y_{i}=0$ in Eq.~\ref{eq:rarmse})
We're trying to understand if \Fathom has a bug related to delivery rates on certain connections.)}

\begin{figure*}[htb!]
     \begin{subfigure}[]{\textwidth}
          \centering
         \includegraphics[width=\textwidth]{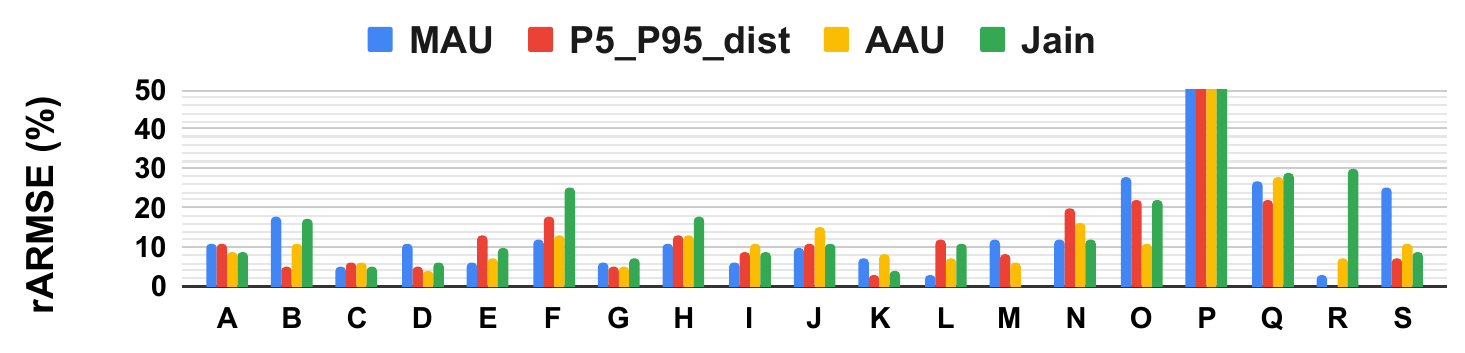}
         \caption{Small RPCs. Some error bars exceed 50\%.}
    \end{subfigure}

    \begin{subfigure}[]{\textwidth}
    \centering
    \includegraphics[width=\textwidth]{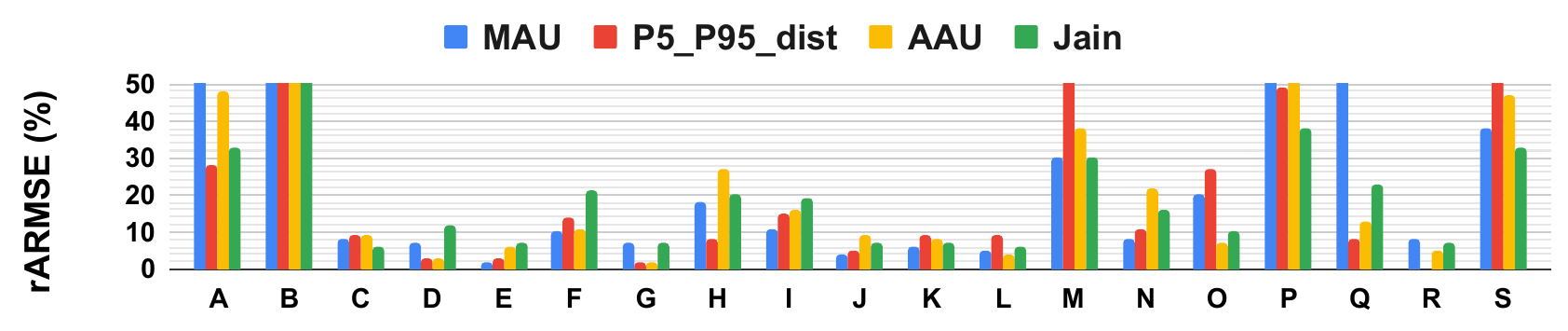}
    \caption{Large RPCs. Some error bars exceed 50\%.}
    \end{subfigure}

  \caption{\small Prediction error for tail (p99) transmit latency, medium QoS.}
  \label{fig:multi-qr95-medqos}
\end{figure*}

\begin{figure*}[htb!]
    %  \begin{subfigure}[]{0.75\textwidth}
          \centering
         \includegraphics[width=\textwidth]{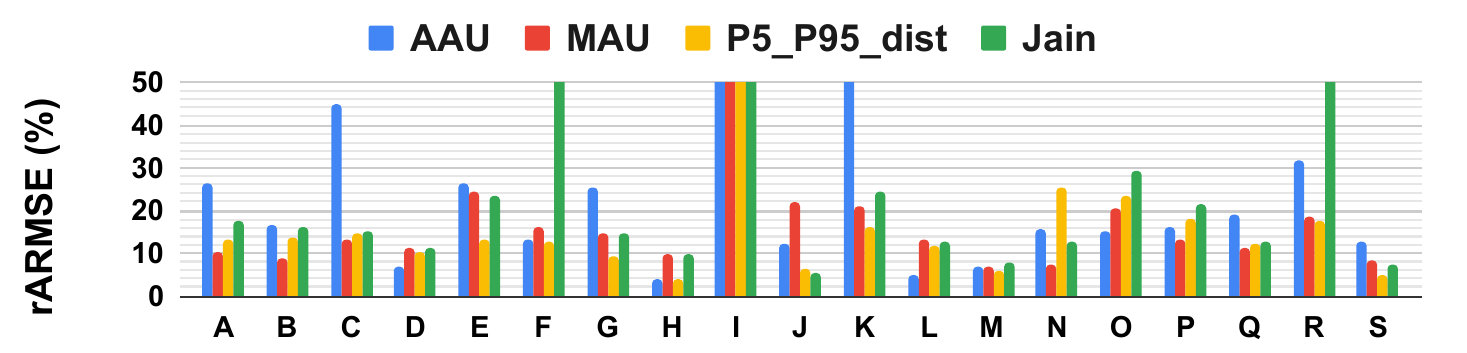}
          \caption{\small P1 delivery-rate prediction error, medium-QoS. Some error bars exceed 50\%.}
        \label{fig:multi-qr95-delivery-rate}
\end{figure*}

% \begin{figure}[htb]
%      \centering
%          \includegraphics[width=0.45\textwidth]{figures/linearvsqueue-qr95.pdf}
%          \\Set of possible \AFMs: \{p99 XXX\}
%         \caption{Number of \AFMs accurately predicted by linear and/or queueing models.}
%         \label{fig:linearvsqueue-qr95}
% \end{figure}

\begin{table*}[htb]
    \centering
    \footnotesize
    \begin{adjustbox}{width=\textwidth}
    \begin{tabular}{l||c|c|c|c|c|c|c|c|c|c|c|c|c|c|c|c|c|c|c|}
             & \multicolumn{19}{c}{Fabric} \\ \cline{2-20}
        \AFM & A & B & C & D & E & F & G & H & I & J & K & L & M & N & O & P & Q & R & S \\ \hline \hline
        p50 latency/1kb &  LQ & LQ & LQ & LQ & LQ & LQ & LQ & LQ & LQ & LQ & LQ & LQ & LQ & LQ & LQ & LQ & LQ & LQ & LQ \\ \hline 
        p99 latency/1kb &  LQ & LQ & LQ & LQ & LQ & LQ & LQ & L  & LQ & X  & X  & X  & X  & LQ & X  & X  & X  & L  & LQ \\ \hline 
        p50 latency/64kb & LQ & LQ & LQ & LQ & LQ & LQ & LQ & LQ & LQ & LQ & LQ & LQ & LQ & LQ & LQ & LQ & LQ & LQ & LQ \\ \hline 
        p99 latency/64kb & LQ & X & LQ  & LQ & X  & X  & LQ & X  & LQ & X  & X  & X  & X  & LQ & X  & X  & X  & X  & X  \\ \hline 
 %        & 
    \end{tabular}
    \end{adjustbox}
    \\ L$=$ predicted by linear model; Q$=$ predicted by queuing-based model; X$=$ predicted by neither.
    \caption{Successful latency predictions (rARMSE$\le 0.15$) of linear and queueing-based models, low QoS.}
    \label{tab:linearvsqueue-qr95-low}
\end{table*}
\subsection{Linear vs. queueing-based models}
\label{sec:linear-or-queueing}

\trf{tab:linearvsqueue-qr95-low} shows the number of tail \AFMs predicted, within our 15\% threshold, by linear and/or queueing-based models, for fabric-wide low-QoS flows.
In every case, linear models make at least as many successful predictions as queuing-based models, and sometimes
more.  Note that \frf{fig:stability-64kbp99} shows that, from week to week, queueing-based models using MAU are sometimes slightly more successful, but \trf{tab:linearvsqueue-qr95-low} covers all possible \NLMs.
\trf{tab:linearvsqueue-qr95-med} in \srf{sec:more-inter-block} shows that for medium-QoS flows, 
linear and queueing-based models were equally successful.

\subsection{Coverage}
\label{sec:coverage}

Our approach first finds a knee in the training data, then constructs a model that covers \NLM values at and
below a knee-based threshold.  This raises the question of what fraction of the domain (x-axis buckets)
of testing-data samples is covered by such
a model, if we do find a high-accuracy model at all.

In most cases, since our networks are often not heavily utilized, we do not detect a knee, and so our models
cover 100\% of the data. 
In cases where there is a knee, since the samples are not uniformly distributed across all 
\NLM values, coverage is not a simple function of the threshold.

For fabric-wide predictions for our set of 19 fabrics, we looked at the best models for p50 and p99 transmit latencies for both
1kb and 64kb RPCs, for both low and medium QoS, and in most cases either the coverage was 100\% or we found no accurate
model.  In just five cases (only for p99 latencies), coverage for an accurate model was below 100\%, ranging from 83\% (fabric $C$, low QoS, 1kB RPCs)
to 99\% (fabric $N$, low QoS, 64kb RPCs).  It is possible that using finer-grained buckets would improve this further.

\section{Sensitivity Analysis}
~\label{sec:sensitivity}
Our method has several parameters that can affect either our predictions, or how we measure
their success.   In this section, we describe the effects of varying some of these parameters.
%In \srf{sec:eval_case_study} we demonstrated how our methodology can be applied to a real-world DCN fabric and make useful predictions.
%We now look at how tuning different knobs/parameters of our methodology affects the characteristics of the prediction results using the same fabric \emph{T}. 

\subsection{Regression Error Threshold}
\label{sec:error-threshold}

Users can define a threshold rARMSE, based on their operational needs and risk tolerance.
A lower threshold reflects a lower tolerance for mis-predictions, but also reduces the number of
cases where our method can make successful predictions at a specified target quantile, such as QR95.

\frf{fig:sensitivity_model_count_low} shows the number of latency-based \AFMs we could predict, with $rARMSE \le 0.15$,
over the 19-fabric dataset, out of eight possible \AFMs (four different RPC sizes, at both p50 and p99), for low-QoS flows.
As one would expect, the number decreases as the threshold tightens.
\frf{fig:sensitivity_model_count_med} shows the same analysis for medium-QoS flows.

\begin{figure}[ht]
  \begin{minipage}{0.49\textwidth}
     \centering
     \includegraphics[width=\linewidth]{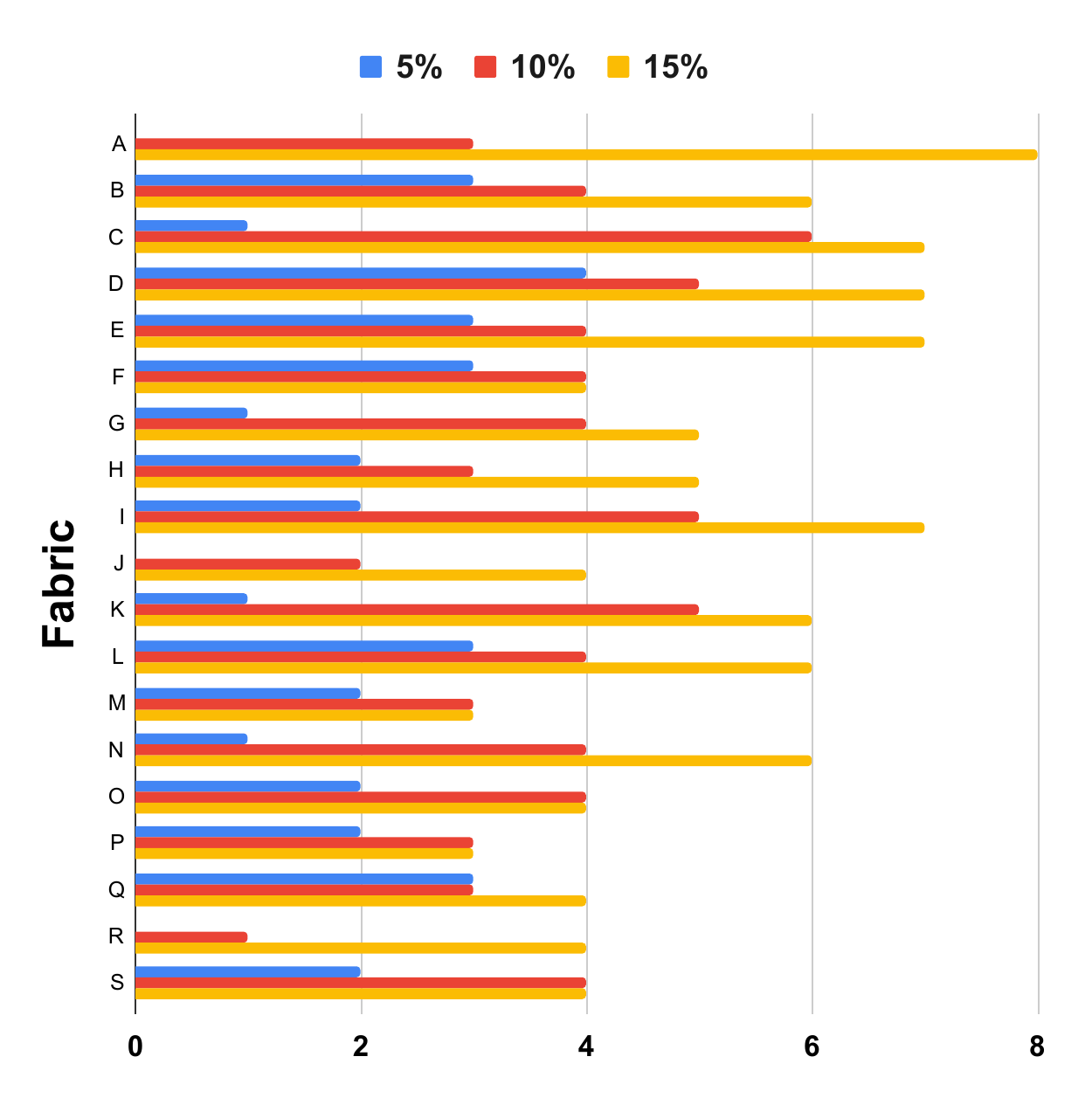}
    \footnotesize Total number of \AFMs analyzed = 8.\\
     \caption{\small{\# of accurately-predictable \AFMs vs. error thresholds, low QoS.}}
     \label{fig:sensitivity_model_count_low}
  \end{minipage}
  \begin{minipage}{0.49\textwidth}
     \centering
     \includegraphics[width=\linewidth]{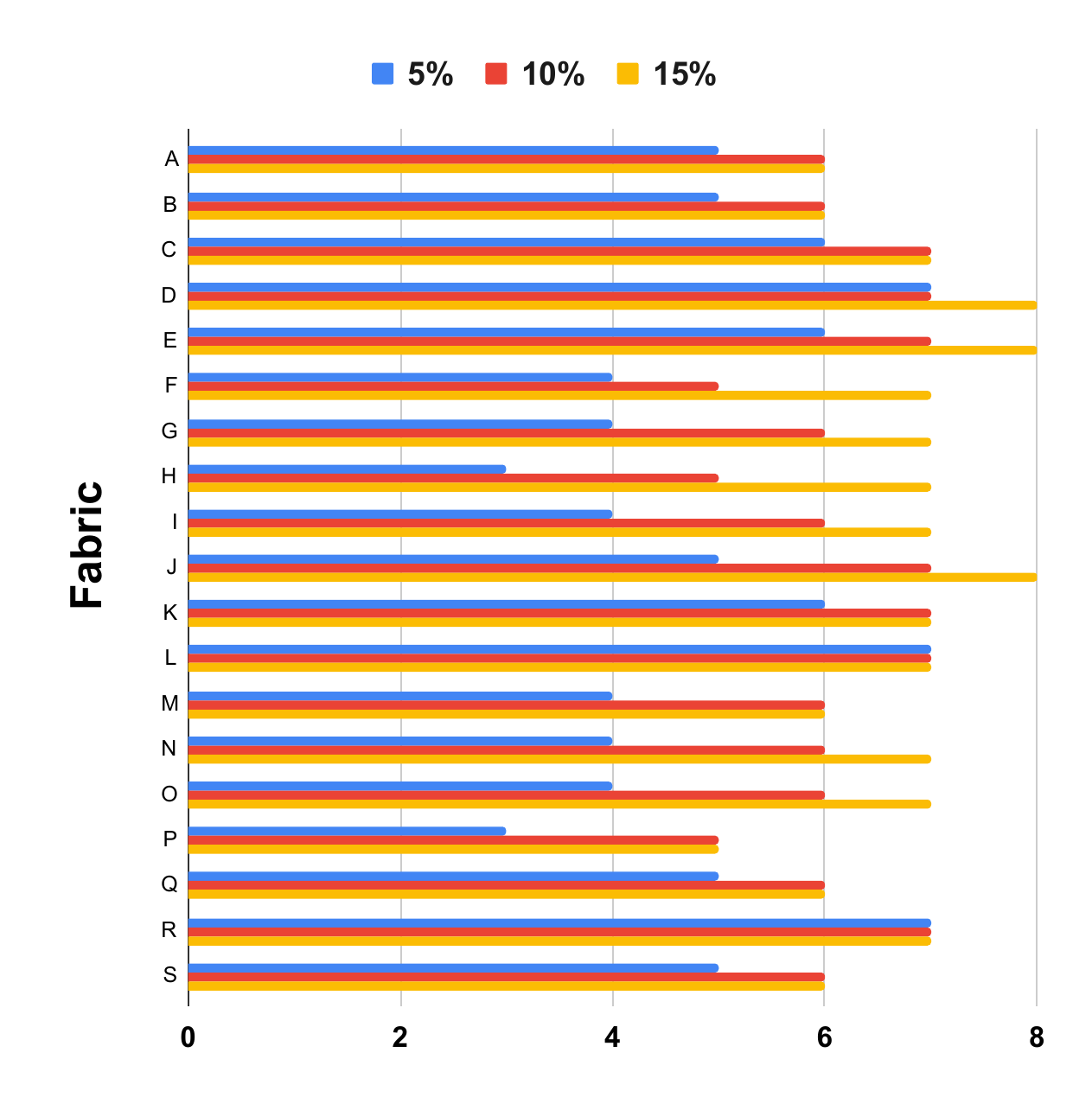}
    \footnotesize Total number of \AFMs analyzed = 8.\\
     \caption{\small{\# of accurately-predictable \AFMs vs. error thresholds, medium QoS.}}
     \label{fig:sensitivity_model_count_med}
  \end{minipage}
\end{figure}

\subsection{Impact of Asymmetric Bias}
\label{sec:sensitivity-bias}
Our definitions of AMSE and rARMSE include an $\alpha$ parameter.
Small values of $\alpha$ penalize underpredictions;
large values penalize overpredictions.
(We expect most users would prefer to overpredict latency, rather than underpredict.)

\begin{figure}[ht!]
    \begin{minipage}[b]{.5\textwidth}
    \centering
    \includegraphics[width=\linewidth]{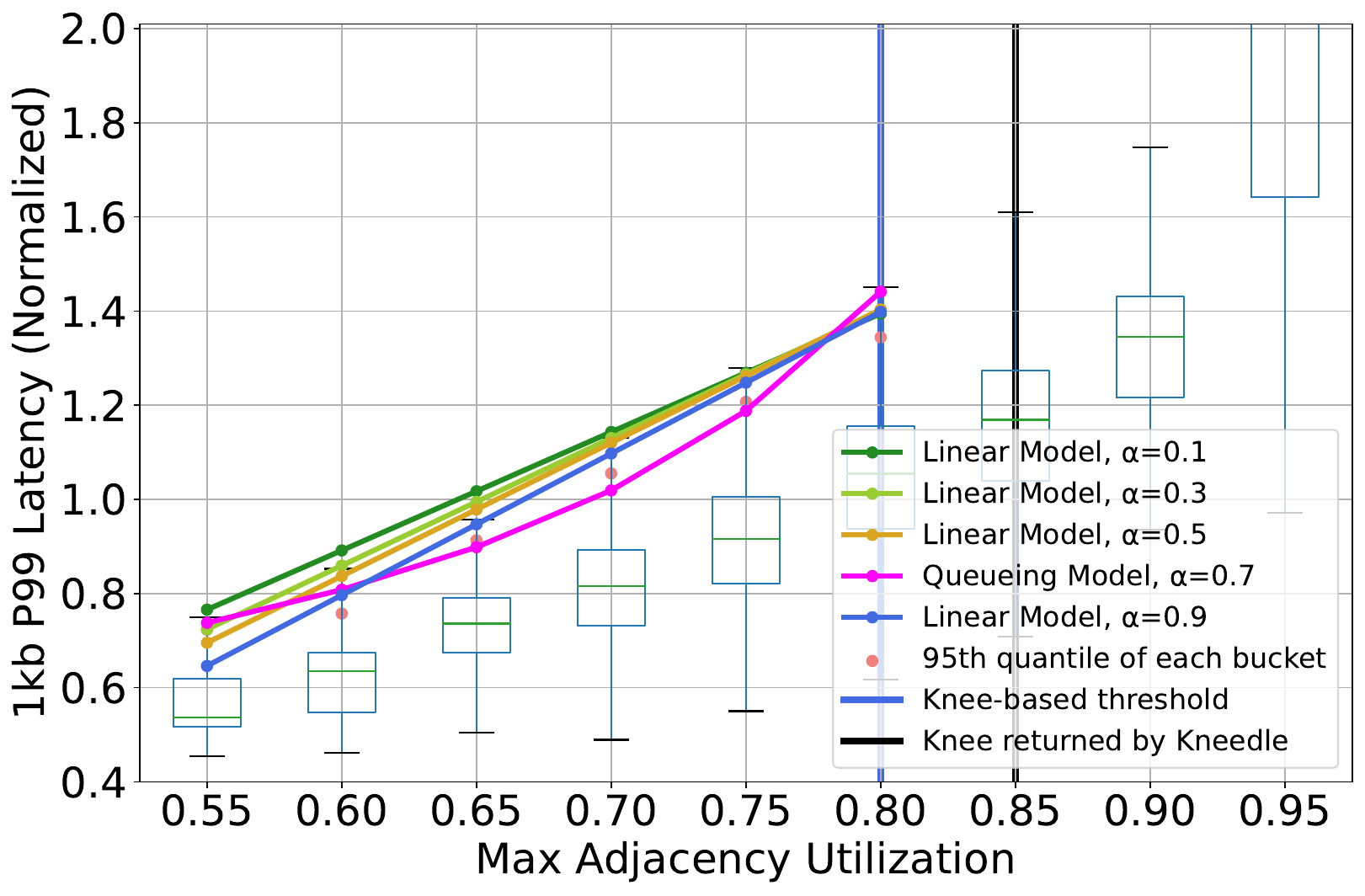}
    \caption{Sensitivity to $\alpha$ for Fabric \emph{C}}
    \label{fig:alpha-fabric-t}
    \end{minipage}
    \begin{minipage}[b]{.5\textwidth}
    \centering
    \includegraphics[width=\linewidth]{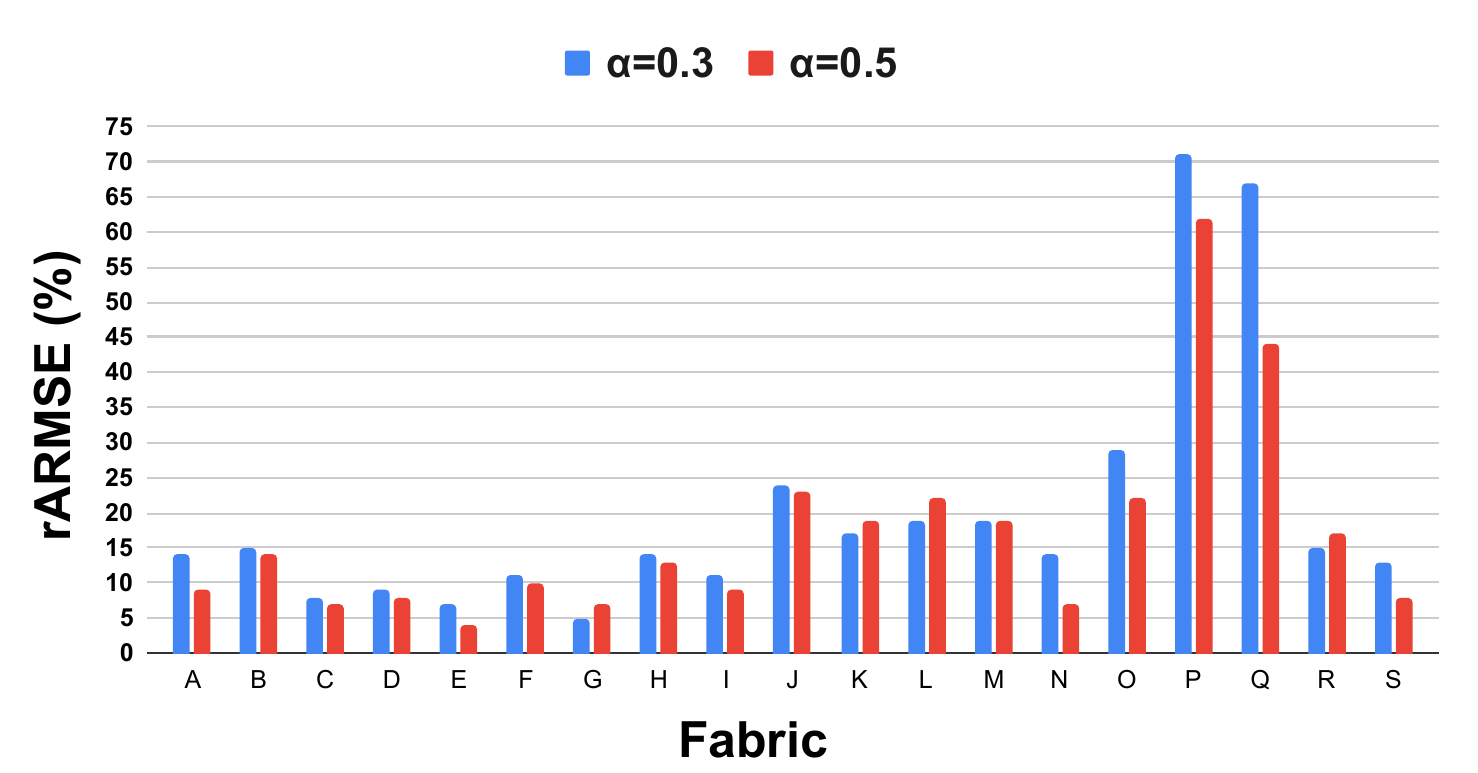}
    \caption{Sensitivity to $\alpha$ across multiple fabrics.}
    \label{fig:alpha-all-fabrics}
    \end{minipage}

\end{figure}

We repeat the single-fabric analysis from \frf{fig:case_study_tail_biweekly}, for p99 (tail)
latencies, low QoS, \NLM$=$MAU, with $\alpha=0.1$, $0.3$, $0.5$ (no bias), $0.7$, and $0.9$.
\frf{fig:alpha-fabric-t} shows that, as expected, as we decrease $alpha$ we bias towards overprediction (SLO preservation),
especially at smaller utilizations.
For $alpha=0.7$, we find that the best model is queueing-based.  We are not sure why; however, the latencies predicted by
the ``best'' and ``worst'' of these models are not significantly
different.

We looked at fabric-wide rARMSE for all 19 fabrics, with $\alpha=0.3$ and $\alpha=0.5$.
\frf{fig:alpha-all-fabrics} shows (for p99 latencies, low QoS,  1kb RPCs, best-predictor \NLMs) that
we get similar accuracies even when the user wants to bias in favor of overprediction.
I.e., if we create a biased model and then use a biased score, the results are close
to what we hoped for, except for fabrics (e.g., $P$ and $Q$) where we get no accurate models.

\subsection{Kneedle Curvature Threshold}
\label{sec:curvature-threshold}

We augmented Kneedle with %an absolute 
curvature threshold $0 < \mathcal{C} < 1.0$,  so that Kneedle only returns knees representing
large changes in the \AFM.
Large $\mathcal{C}$ values reduce false-positive knees, and might make the knee position more stable w.r.t. changes
in input data, but risk non-detection of actual knees.

\begin{table}[htb]
\small
\begin{tabular}{|c|c|c|c|c|c|c|}
\hline
\textbf{\AFM / Threshold $\mathcal{C}$} & \textbf{0.05} & \textbf{0.1} & \textbf{0.3} & \textbf{0.5} & \textbf{0.7} & \textbf{0.8}  \\ \hline
% \textbf{1kb Median Tx. Latency}   & MAU              & MAU                 & MAU               \\ \hline
% \textbf{64kb Median Tx. Latency}  & MAU              & P5\_P95\_Dist       & X                 \\ \hline
\textbf{1kb p99 Tx. latency}     & 0.85 & 0.85 & 0.85 & 0.85 & 0.85 & X              \\ \hline
\textbf{64kb p99 Tx. latency}    & 0.9 & 0.9 & 0.9 & 0.9 & 0.9 & X                  \\ \hline
\end{tabular}
\\Fabric \emph{C}, low-QoS RPCs
\caption{\small A stable example of knee values vs. curvature thresholds.}
\label{tab:curvature_1kbp99}
\end{table}

\begin{figure}[htb]
    \centering
    \small
    \includegraphics[scale=0.45]{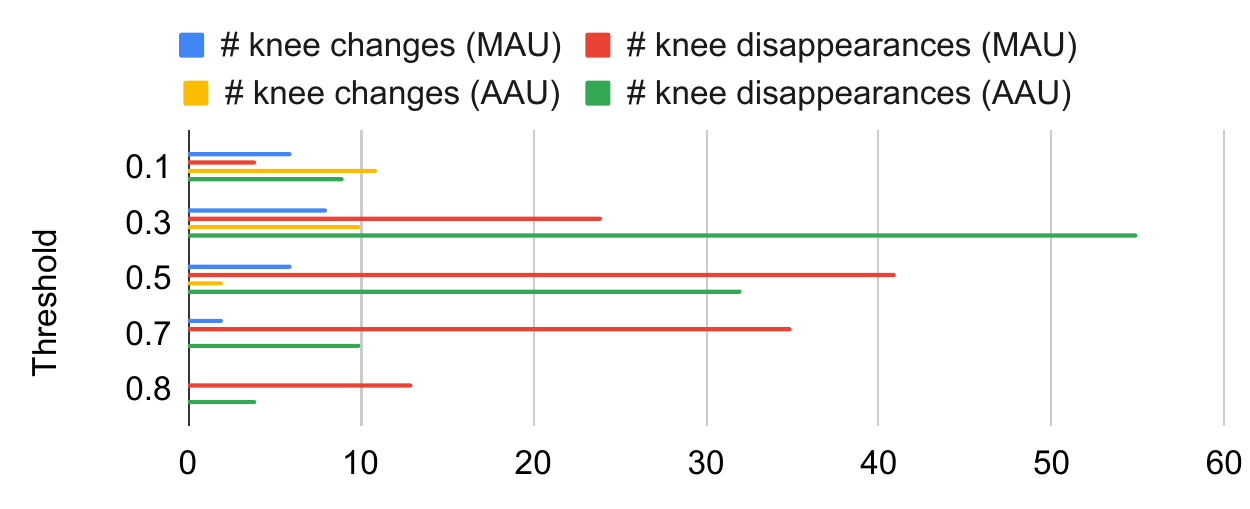}
    \caption{Effect of threshold  $\mathcal{C}$ on knee-finding}
    \label{fig:knees-vs-thresholds}
\end{figure}

Our expectations were that the choice of $\mathcal{C}$ would have limited impact on whether and where we detect knees. 
This is true for some combinations of fabric, QoS, and \AFM (see
\trf{tab:curvature_1kbp99}). 
In other cases, knees are sensitive to $\mathcal{C}$, and with no fixed pattern.  \frf{fig:knees-vs-thresholds} summarizes results over all 1520 combinations of fabric, QoS, \AFM, and threshold (above the minimum of 0.05).  It shows that increasing the threshold causes some knee positions to change, and higher thresholds do cause knees to disappear.  (We are not sure which of
these are ``false'' knees.) 
Note that in 82\% of the cases (1247/1520), threshold changes had no effect, but for now we need users to experimentally validate their choice of threshold.
\section{Discussion}
\label{sec:takeaways}

\subsection{Takeaways}
We summarize several takeaways.  
These are generalizations to our limited data set, and may not always hold true. 
% These are generalizations, and even in our limited data set, they do not always hold true. 
% Some of these are expected (i.e., confirm
% our intuition or conventional wisdom); 
Some confirm our intuition or conventional wisdom, some are possibly surprising.

\parab{We can construct models that predict \AFMs based on individual \NLMs}, especially at the fabric level and less
often at the block level.

\parab{No single \NLM is universally the best predictor.}
Our multi-fabric analysis shows that in most cases, the best predictor \NLM for an \AFM depends on the \AFM, QoS, and fabric.
Much prior work \arxivedit{(e.g., ~\cite{HeEtAl2007,KandulaEtAl2005,WangEtAl2006, jupiter-evolving})} has focused on using MAU or MLU as a predictor; our results suggest that focus is too narrow, as these \NLMs are not always the best predictor for \AFM performance.
\highlightedit{
Also, because different NLMs predict different knees depending on fabric, it might be safest to use the worst-case AFM-NLM relationship to find the danger zone, rather than hoping for a best-single NLM.
}

\parab{For most cases we studied, a linear relationship fits the data well.} Conventional wisdom
in our organization was that at low utilizations, RPC latencies would not depend much on \NLMs, and if
there was a dependency, a queueing-based model would be the best fit.   In fact, we found that the best
model is almost always linear, even at low utilizations.

\parab{Prediction success depends on QoS.}  
This confirms our expectations, as the goal of QoS mechanisms is to differentially protect higher-QoS flows from network conditions.

% Not surprising, since the goal of QoS mechanisms is to differentially protect
% higher-QoS flows from network conditions.

\parab{Short-message latencies are more predictable than long-message latencies.} 
This adds weight to conventional wisdom that congestion-control algorithms tend to respond too slowly to network conditions to affect
the sender's behavior for a short (e.g., 1-packet) message.

\parab{Median latencies are easier to predict than tail latencies.}  This matches our expectations, since generally
tail behavior is the result of more complex network behavior.

%\vspace{0.01in}\noindent \textbf{Prediction accuracy varies somewhat from from week to week,}
%\textbf{but the nature of a relationship} %(linear vs. queueing-based) is stable.
\parab{Prediction accuracy varies from from week to week}, even when we re-train models
every two weeks. 
This suggests that, in at least some fabrics, workloads vary enough to make a static model insufficient. 
However, the nature of a relationship (linear vs. queueing-based) appears to be more stable.

% \subsection{Implications for stakeholders}
% \textcolor{orange}{
% Based on our takeaways, we provide several suggestions and promising future directions for \NLM-\AFM relationship stakeholders. 
% }

\subsection{Limitations of our approach}
\label{sec:limitations}

Our work has several limitations due to constraints of the measurement infrastructure and environment.
Not all of the traffic in our network is RPC-based, and not all of the RPC-based traffic is instrumented.
We are currently aggregating data into 5-minute buckets, due to \Fathom's limitations, and this hides
the effects of short-term bursts.  (An alternative would be to collect a burst metric such as the
maximum instantaneous queue depth during a measurement period, but our switch instrumentation does
not provide this.)

Our results might not be representative for \highlightedit{all types of application mixes and fabric types}
due to several reasons, including the nature of our specific workloads,
and the design of our networks and their topologies.   
In particular, we gathered traces from fabrics
which are not heavily used by recently-appearing machine-learning workloads, such as training for large
language models.

\Urlmuskip=0mu plus 1mu\relax

\bibliographystyle{ACM-Reference-Format}
\bibliography{reference}

\appendix

\section{Ethics}

This work does not raise any ethical issues.  By design, we did not
collect any user-identifying data or user data.

\section{Details of latency-\AFM measurement}
\label{fathom-timeline-details}

\frf{fig:fathom-timeline} illustrates how \Fathom measures RPC ``transmit latency.''  Note
that the time for the last measured byte of a message to transit the network is not included
in this measurement.

\begin{figure}[htb]
     \centering
         \includegraphics[width=\linewidth]{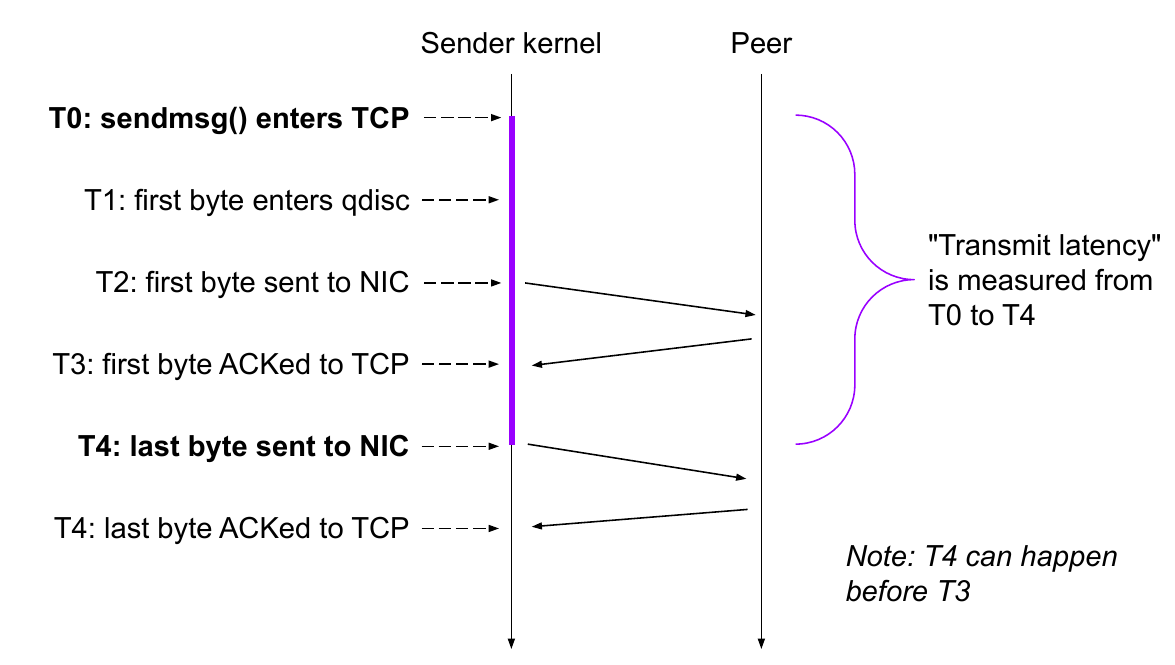}\\
         \caption{Timing for \Fathom transmit latency measurement}
        \label{fig:fathom-timeline}
\end{figure}

\section{Effects of quantile-regression thresholds}
\label{sec:qr-thresholds}

In \srf{ssec:quantile_regression} we described quantile regression, and wrote that we focused on 95th quantiles
because at higher percentiles, our datasets were too sparse in some cases to compute low-noise values.)    Here
we compare how our ability to predict 99th quantile (``QR99'') \AFMs compares with our ability to predict
QR95 \AFMs.

\begin{figure*}[htb]
     \centering
     \begin{subfigure}[]{0.5\textwidth}
         \centering
         \includegraphics[width=\textwidth]{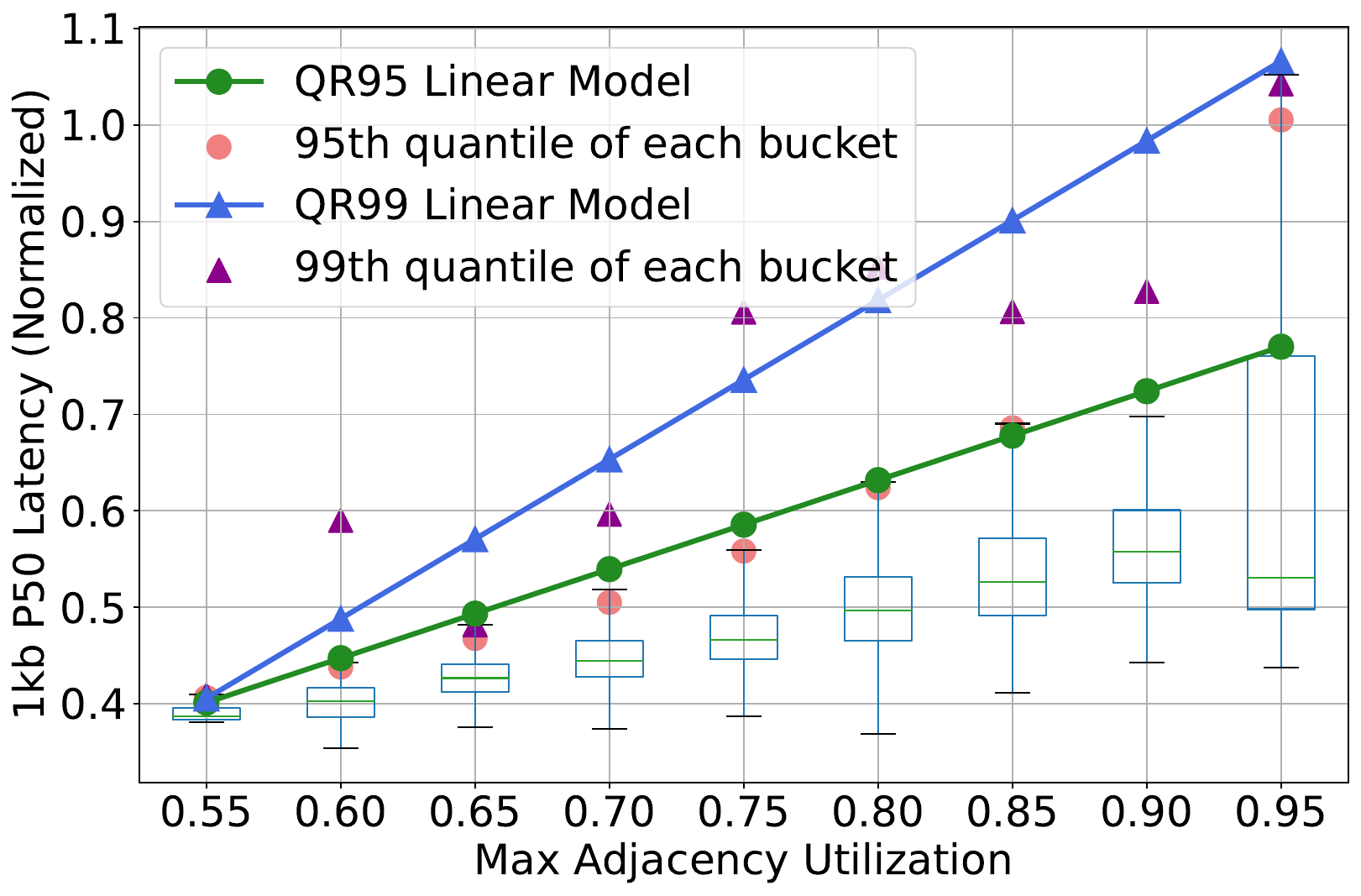}
         \caption{1kb Median Transmit Latency.}
     \end{subfigure}%
    \hfill
     \begin{subfigure}[]{0.5\textwidth}
         \centering
         \includegraphics[width=\textwidth]{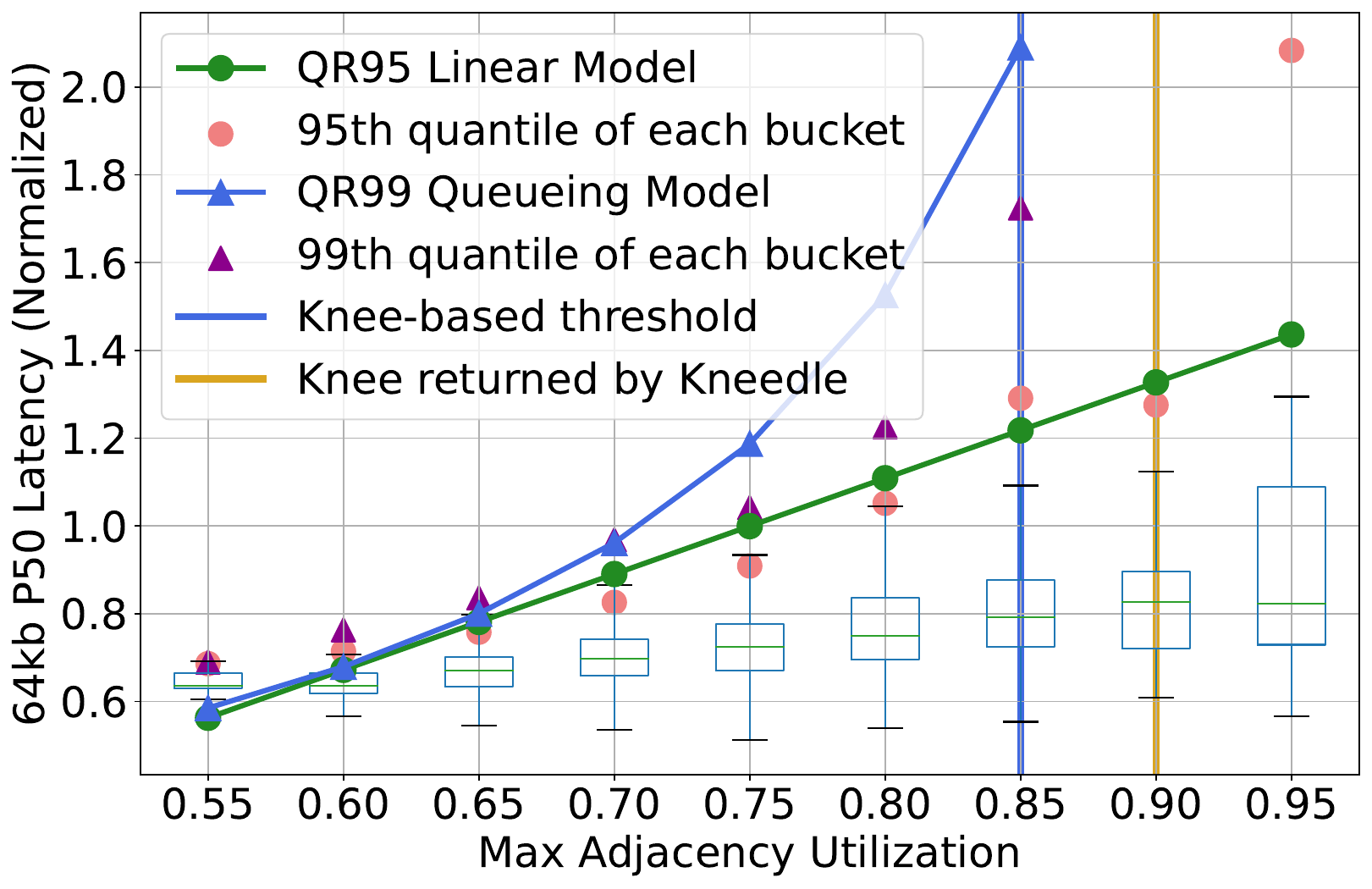}
         \caption{64kb Median Transmit Latency.}
     \end{subfigure}
     \\ Fabric \emph{C}, low QoS, \NLM$=$ MAU
    \caption{Comparing QR95 and QR99 models for median latency}
        \label{fig:case_study_median_biweekly-qr99}
\end{figure*}

\begin{figure*}[htb]
     \centering
     \begin{subfigure}[]{0.5\textwidth}
         \centering
         \includegraphics[width=\textwidth]{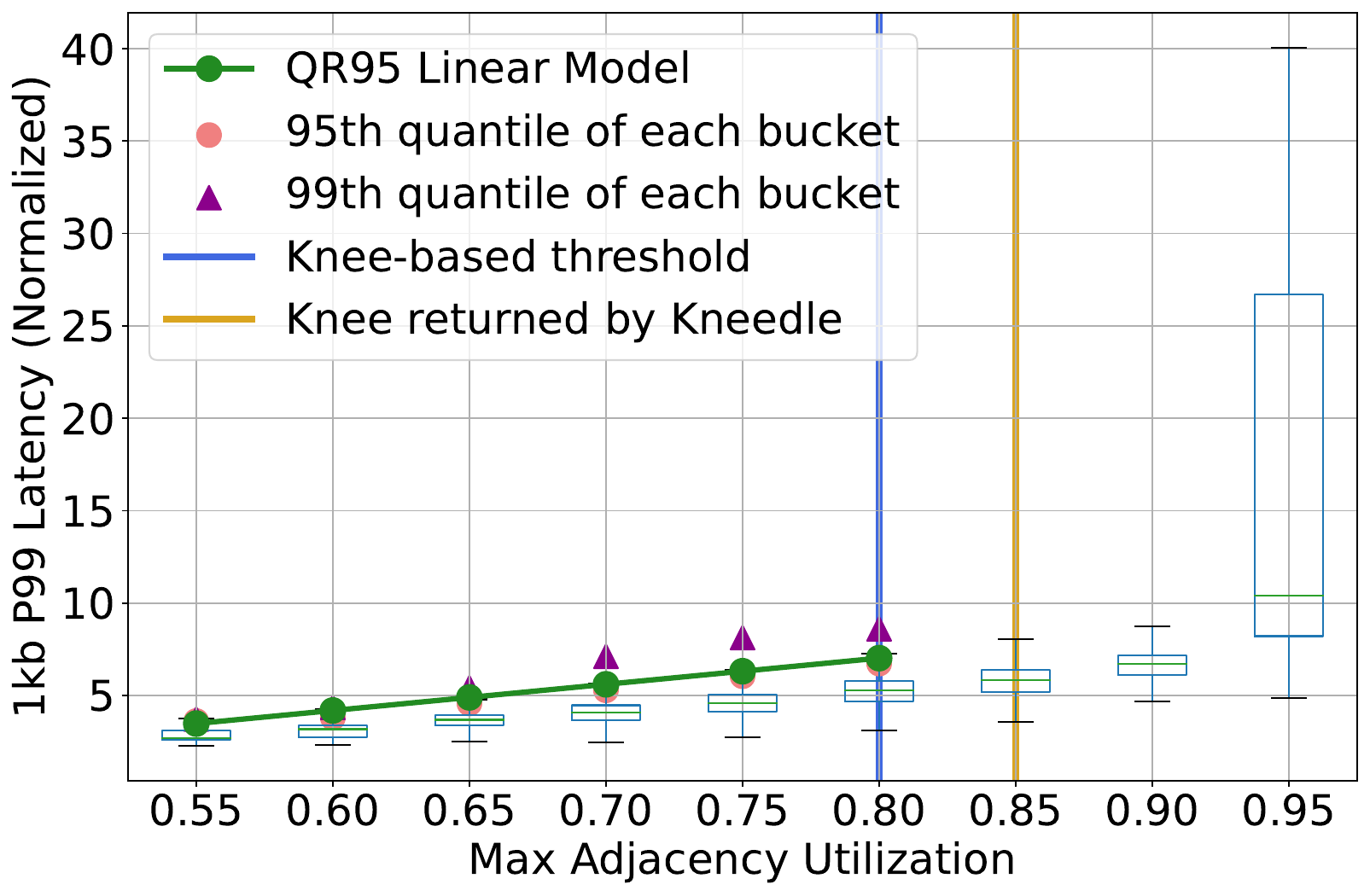}
         \caption{1kb Tail Transmit Latency.}
         \label{sfig:1kbp99lowboxplot-qr99}
     \end{subfigure}%
     \hfill
     \begin{subfigure}[]{0.5\textwidth}
         \centering
         \includegraphics[width=\textwidth]{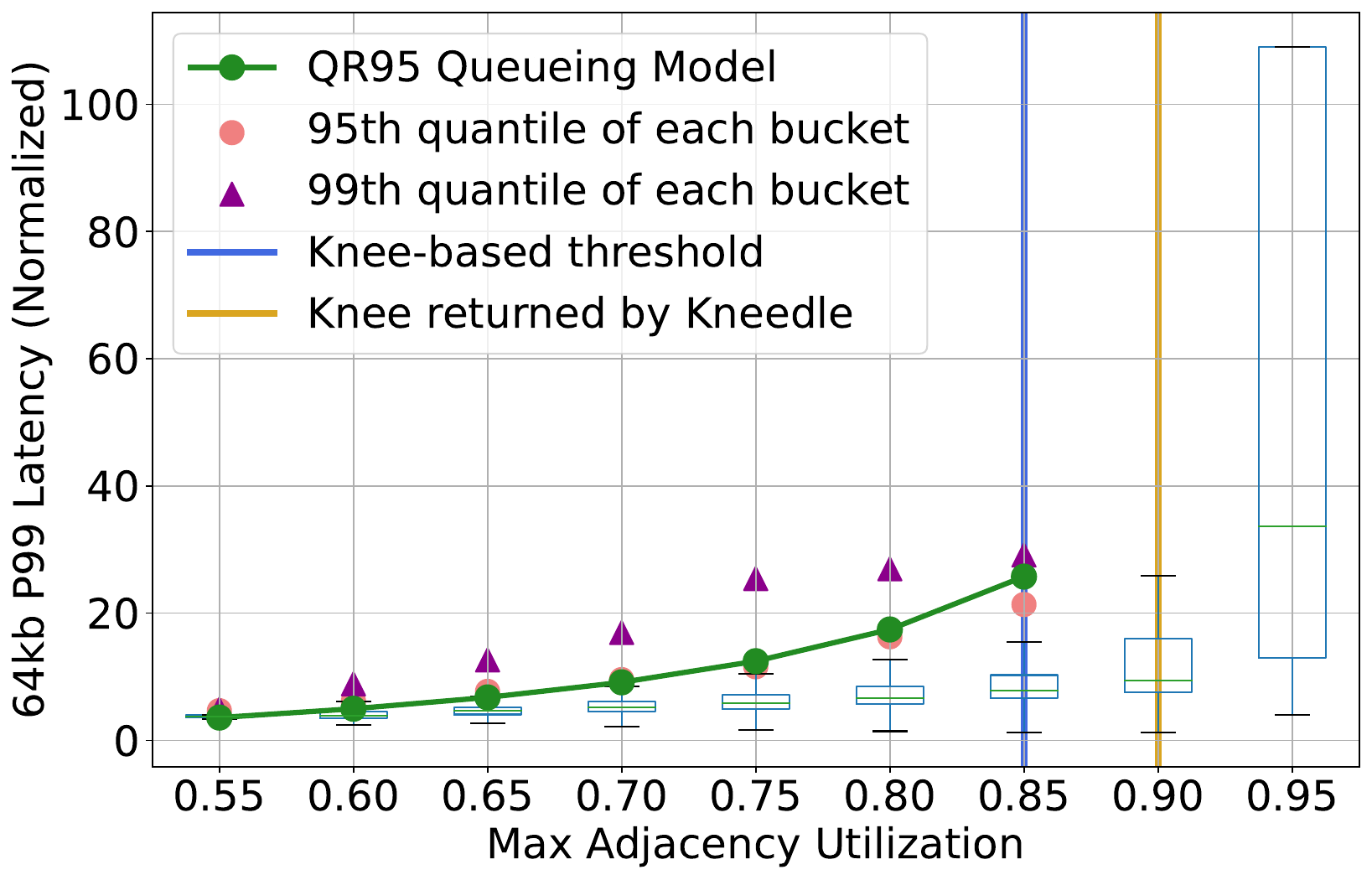}
         \caption{64kb Tail Transmit Latency.}
     \end{subfigure}
     \\ Fabric \emph{C}, low QoS, \NLM$=$ MAU
              \caption{Comparing QR95 and QR99 models for tail latency}

        \label{fig:case_study_tail_biweekly-qr99}
\end{figure*}

For a specific example,
we looked at fabric-level predictions for Fabric \emph{C}, as we did in \srf{sec:eval_case_study}.
Figures~\ref{fig:case_study_median_biweekly-qr99} and \ref{fig:case_study_tail_biweekly-qr99} show, respectively,
median (p50) and tail (p99) results for low-QoS flows and small-RPC tail-transmit latencies.   The figures show
both 95th and 99th quantile \AFMs for each x-axis bucket, and the QR95 models we obtained in each case.
For median transmit latencies (\frf{fig:case_study_median_biweekly-qr99}), our method did find high-confidence
QR99 models.   However, for tail transmit latencies (\frf{fig:case_study_tail_biweekly-qr99}), we did not
find QR99 models; an examination of the ``99th quantile'' points in that figure suggests that if there is
a curve that would fit these points, it is consistent with neither a linear or queueing-based model.

We are unsure whether this failure to produce high-confidence QR99 models is due to intrinsically different
behavior for the highest-valued samples in our dataset, or just due to higher noise levels in that subset
of the data.

%show the high-accuracy models, if any, under a 95th percentile threshold for quantile regression (\ie QR95 models), for low-QoS flows.
%For tail transmit latencies, we were not able to find any high-accuracy models. 
%While we were able to collect enough training data to perform regression, the 99th quantiles are noisier than 95th quantiles, which presumably leads to less stable results. 
% \fixthis{add a figure here showing an example of how different quantile-regression thresholds lead
% to different models.  If we can provide a QR99 version, great -- otherwise use QR50/QR75/QR95}

\section{Additional inter-block results}
\label{sec:more-inter-block}

\begin{figure*}[t!]
     \begin{subfigure}[]{\textwidth}
          \centering
         \includegraphics[width=\textwidth]{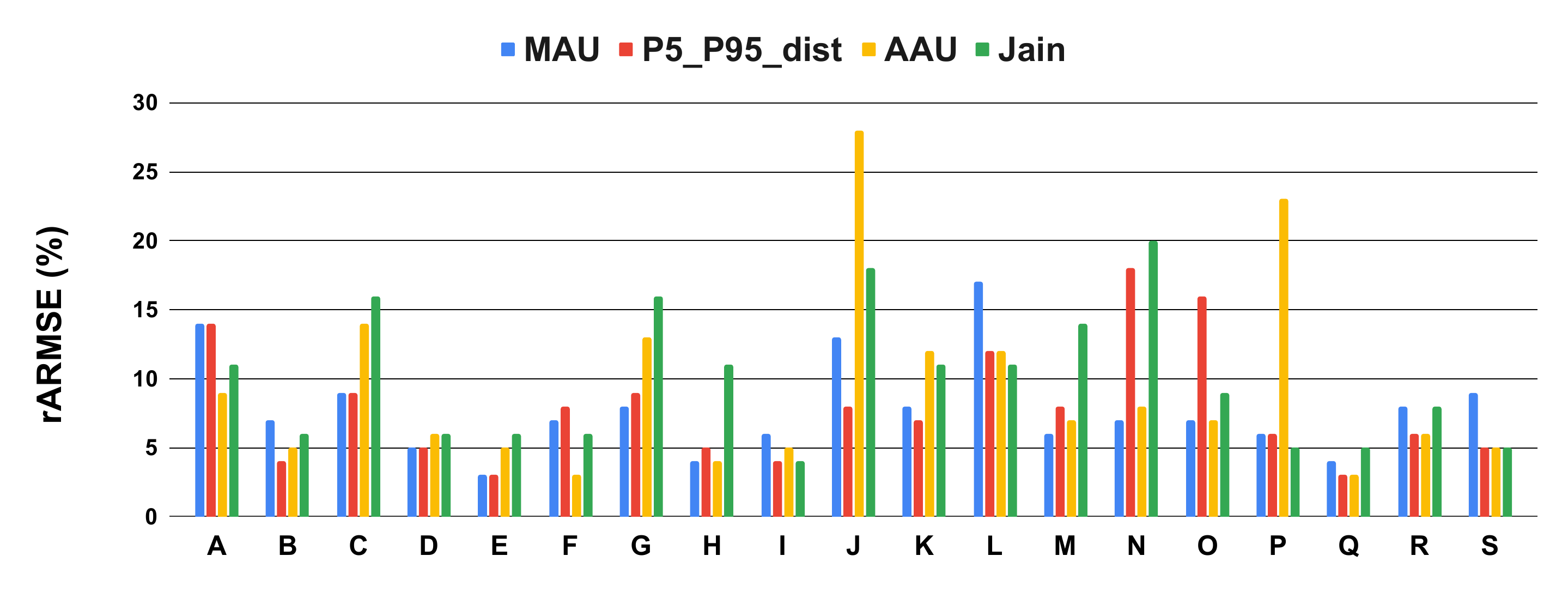}
         \caption{Small RPCs.}
     \end{subfigure}
     \begin{subfigure}[]{\textwidth}
          \centering
         \includegraphics[width=\textwidth]{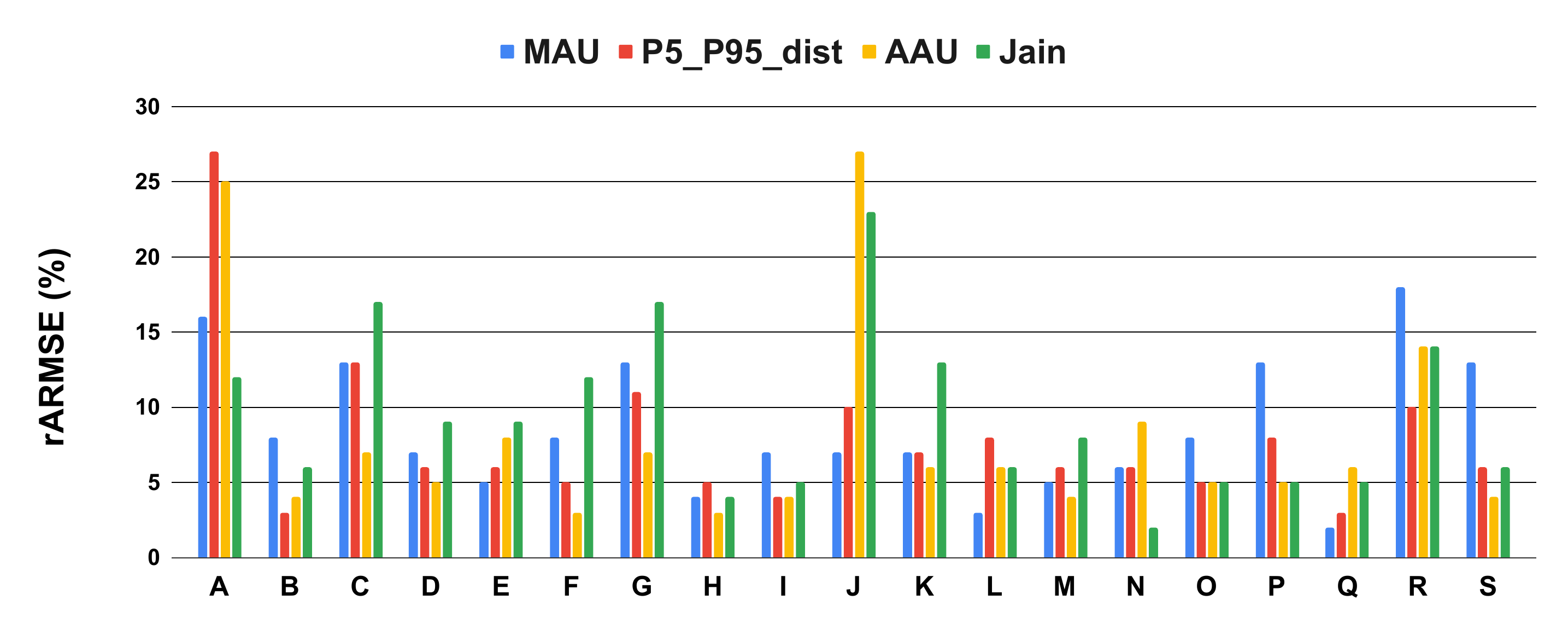}
         \caption{Large RPCs.}
     \end{subfigure}
          \caption{Prediction error for various \AFMs, median (p50) RPC transmit latency, low QoS.}
        \label{fig:multi-qr95-error-median-low}
\end{figure*}

\begin{figure*}[t!]
     \begin{subfigure}[]{\textwidth}
          \centering
         \includegraphics[width=\textwidth]{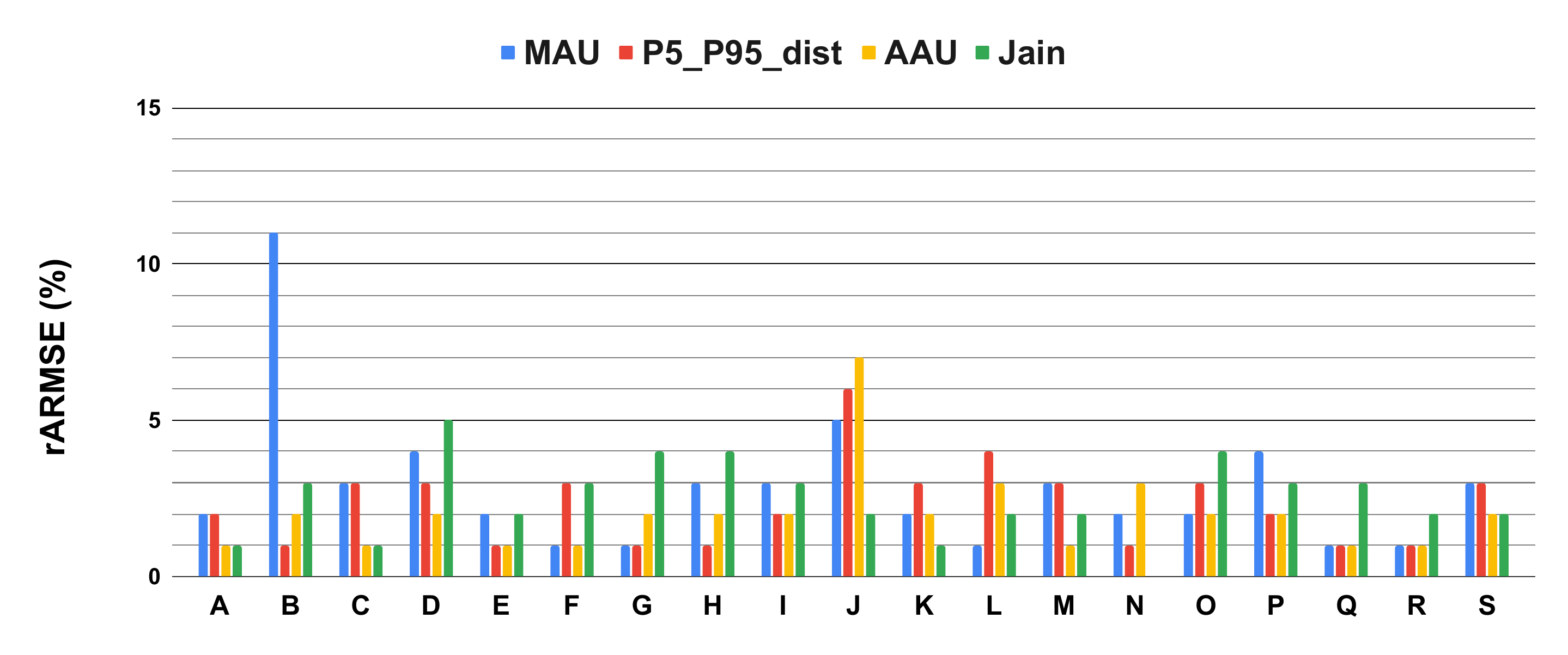}
         \caption{Small RPCs.}
     \end{subfigure}
     \begin{subfigure}[]{\textwidth}
          \centering
         \includegraphics[width=\textwidth]{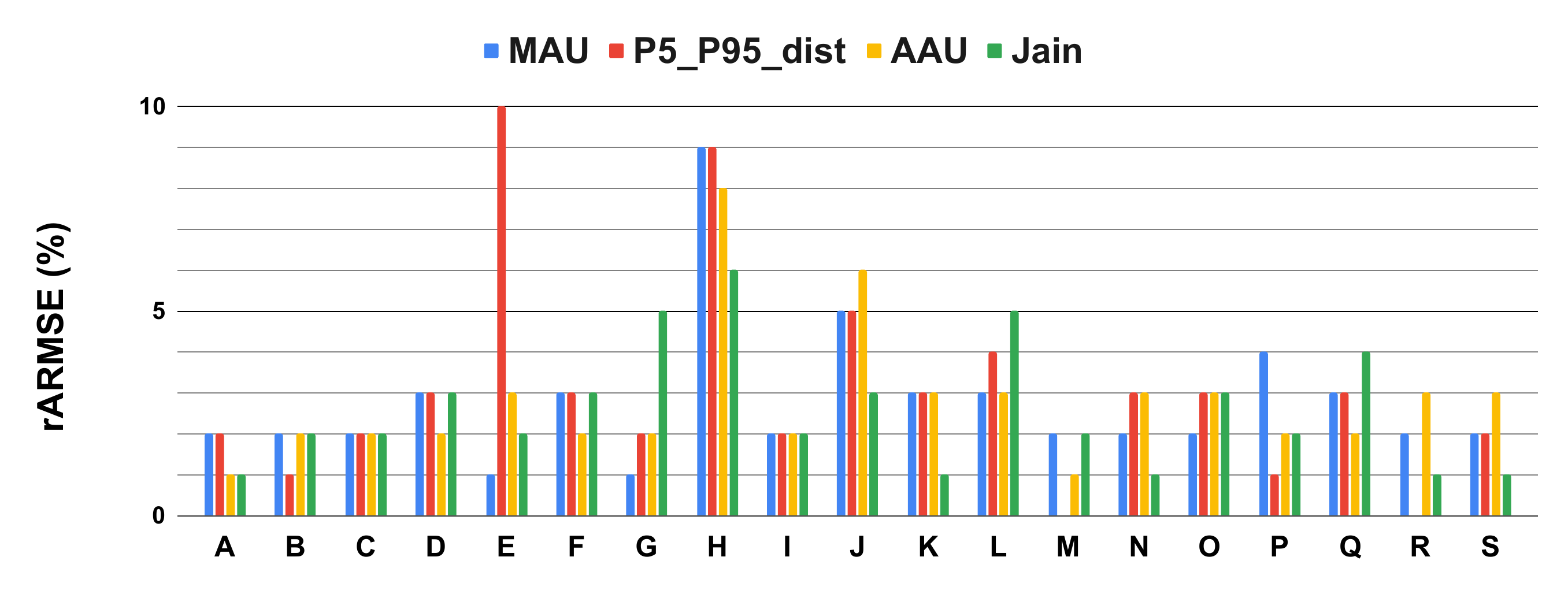}
         \caption{Large RPCs.}
     \end{subfigure}
          \caption{\small Prediction error for various \AFMs, median (p50) RPC transmit latency, medium QoS.}
        \label{fig:multi-qr95-error-median-med}
\end{figure*}

To complement \frf{fig:multi-qr95-error-tail-low} and \frf{fig:multi-qr95-medqos}, which shows per-fabric prediction errors for the p99 (tail) latency of both QoS classes,
Figures \ref{fig:multi-qr95-error-median-low} and \ref{fig:multi-qr95-error-median-med} present p50 (median) latencies for low-QoS and medium-QoS flows, respectively.
Figures \ref{fig:multi-qr95-error-median-low} and \ref{fig:multi-qr95-error-median-med} show that
p50 latencies are easier to predict successfully than p99 latencies.
These results match our expectations.
However, even
for a given fabric, different (QoS, \AFM) combinations can have very different accuracies.   For example,
fabric $\mathcal(J)$ is hard to predict for low-QoS p99 latencies, but apparently much easier to predict
for medium-QoS p99 latencies.   This could be because low-QoS and medium-QoS flows are drawn from significantly
different applications.

\trf{tab:linearvsqueue-qr95-low}, which showed for low-QoS flows that if we could
obtain any ``accurate'' model for a fabric and \AFM, then we could always find an accurate
linear model, but in some cases we could not find an accurate queueing-based model.
\trf{tab:linearvsqueue-qr95-med} shows corresponding results for medium-QoS flows; for these
flows, we can find accurate models for a larger fraction of (fabric, \AFM) combinations,
but here we found accurate queueing-based models in every case where we found an accurate linear model,
and vice versa.

\begin{table*}[htb]
    \centering
    \footnotesize
    \begin{adjustbox}{width=\textwidth}
    \begin{tabular}{l||c|c|c|c|c|c|c|c|c|c|c|c|c|c|c|c|c|c|c|}
             & \multicolumn{19}{c}{Fabric} \\ \cline{2-20}
        \AFM & A & B & C & D & E & F & G & H & I & J & K & L & M & N & O & P & Q & R & S \\ \hline \hline
        p50 latency/1kb &  LQ & LQ & LQ & LQ & LQ & LQ & LQ & LQ & LQ & LQ & LQ & LQ & LQ & LQ & LQ & LQ & LQ & LQ & LQ \\ \hline 
        p99 latency/1kb &  LQ & LQ & LQ & LQ & LQ & LQ & LQ & LQ & LQ & LQ & LQ & LQ & LQ & LQ & LQ & X  & X  & LQ & LQ \\ \hline 
        p50 latency/64kb & LQ & LQ & LQ & LQ & LQ & LQ & LQ & LQ & LQ & LQ & LQ & LQ & LQ & LQ & LQ & LQ & LQ & LQ & LQ \\ \hline 
        p99 latency/64kb & LQ & X  & LQ & LQ & LQ & LQ & LQ & LQ & LQ & LQ & LQ & LQ & X  & LQ & LQ & X  & LQ & LQ & X  \\ \hline 
 %        & 
    \end{tabular}
    \end{adjustbox}
    \\ L$=$ predicted by linear model; Q$=$ predicted by queuing-based model; X$=$ predicted by neither.
    \caption{Successful latency predictions (rARMSE$\le 0.15$) of linear and queueing-based models, medium QoS.}
    \label{tab:linearvsqueue-qr95-med}
\end{table*}

\begin{figure*}[t!]
     \begin{subfigure}[]{\textwidth}
          \centering
         \includegraphics[width=\textwidth]{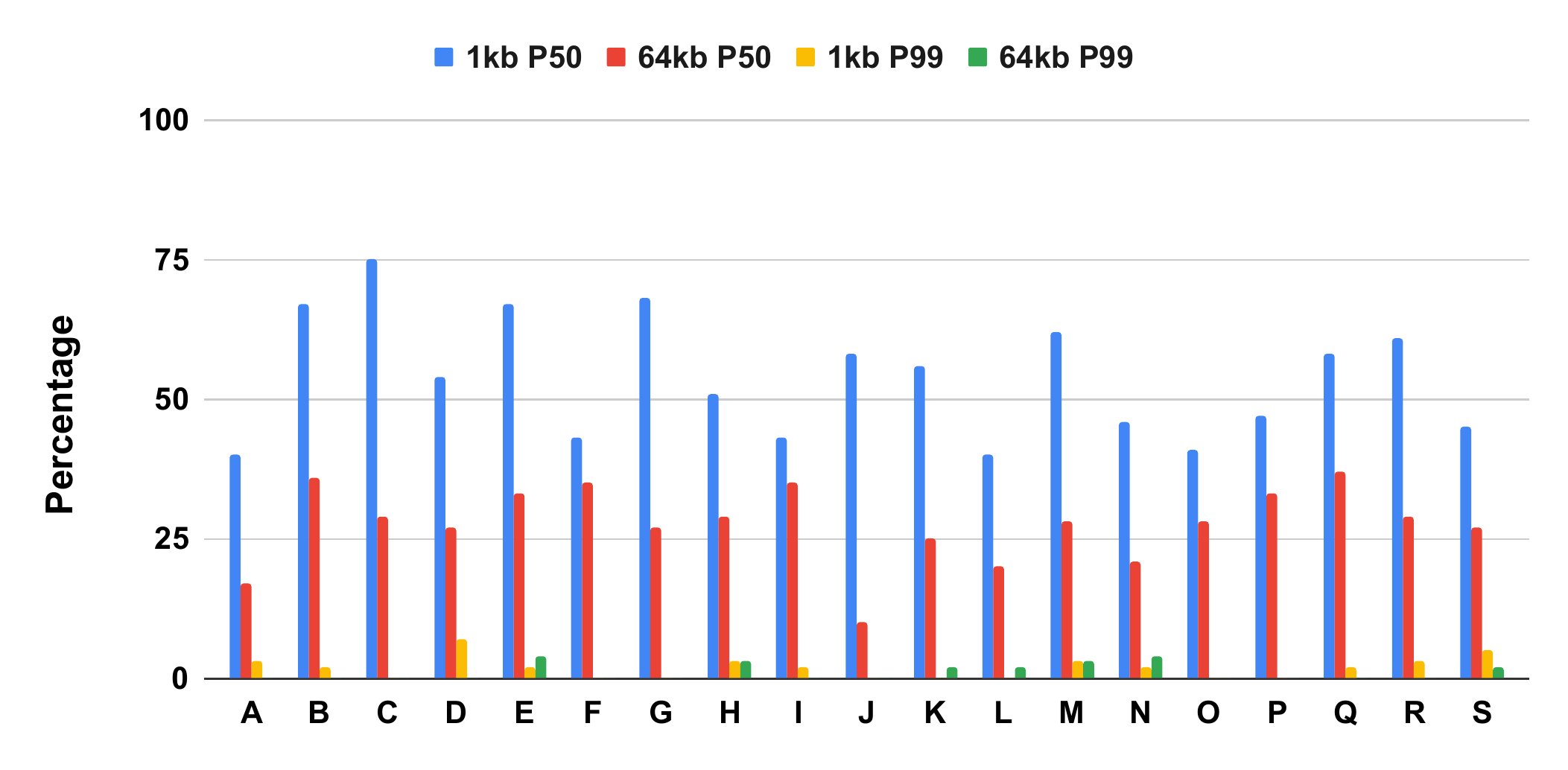}
         \caption{Low QoS.}
         \label{fig:multi-qr95-error-intra-low}
     \end{subfigure}
     \begin{subfigure}[]{\textwidth}
          \centering
         \includegraphics[width=\textwidth]{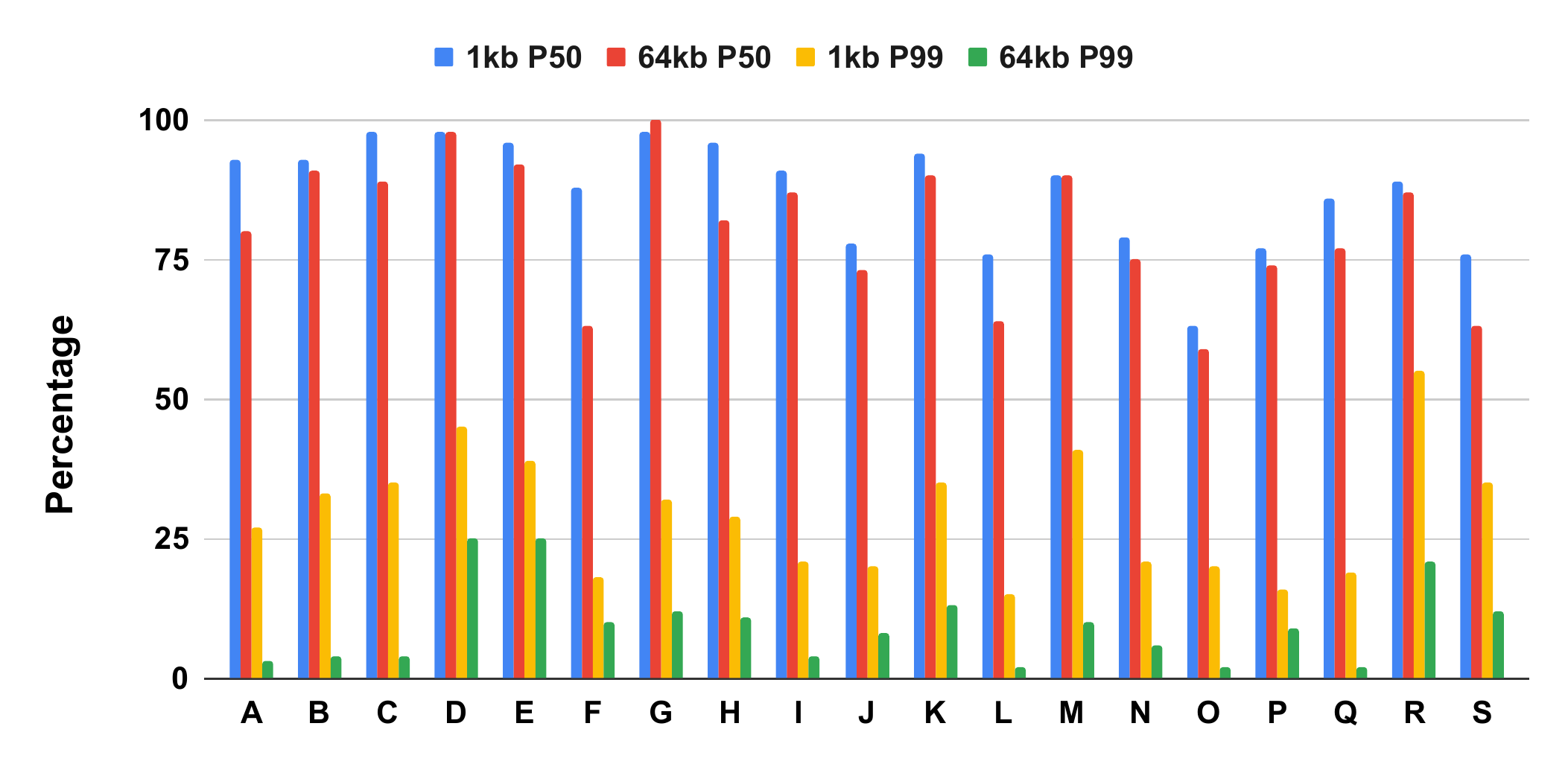}
         \caption{Medium QoS.}
         \label{fig:multi-qr95-error-intra-med}
     \end{subfigure}
          \caption{\small Percentage of aggregation blocks, per fabric, with high-accuracy block-local models.}
        \label{fig:multi-qr95-error-intra}
\end{figure*}

\section{Block-local \NLM Prediction accuracy}
\label{sec:per-fabric-block-level}

In \srf{sec:intra-block-eval}, we looked at how well we could predict \AFMs for RPCs that stay within aggregation
blocks, using \NLMs from non-aggregated intra-block links, for Fabric \emph{C}.   Here we extend that analysis
to all 19 fabrics.

Figure~\ref{fig:multi-qr95-error-intra} shows the percentage of aggregation blocks, per fabric, for which
we obtained at least one high-accuracy model (using any \NLM), broken down by \AFM.
\frf{fig:multi-qr95-error-intra-low} shows that for low-QoS RPCs, we can predict intra-block median latencies
for short RPCs in roughly half of the blocks, and for long RPCs in roughly a quarter of the blocks.
\frf{fig:multi-qr95-error-intra-med} shows that for medium-QoS RPCs, which are generally easier to prodict,
we can predict intra-block median latencies in well over half the blocks (sometimes in almost all of them),
for both RPCs sizes in the figure.
However, we cannot predict tail RPC latencies nearly as well, especially for low-QoS RPCs.

\end{document}